\documentclass[apl,reprint,superscriptaddress,longbibliography]{revtex4-2}
\usepackage{gensymb}
\usepackage{notes2bib}
\usepackage{graphicx}
\usepackage{textcomp}
\usepackage{verbatim}
\usepackage{amsmath}
\usepackage{soul}
\usepackage{setspace}
\usepackage{multirow}
\usepackage{float}
\usepackage[dvipsnames]{xcolor}
\usepackage[colorlinks=true,urlcolor=blue,linkcolor=blue,citecolor=blue,a4paper]{hyperref}
\usepackage{booktabs}
\makeatletter
\def\@fnsymbol#1{\ensuremath{\ifcase#1\or *\or \ddagger\or \dagger\or
   \mathsection\or \mathparagraph\or \|\or **\or \ddagger\ddagger
   \or \dagger\dagger\else\@ctrerr\fi}}
\makeatother
\usepackage[colorlinks=true,urlcolor=blue,linkcolor=blue,citecolor=blue,a4paper]{hyperref}
\graphicspath{{./images/}}

\makeatletter
\renewcommand\p@subsection{}
\makeatother
\begin{document}

\title{\textit{In Situ} Thermal Preparation by CO$_2$-Laser and Chemical Characterization of Oxide Substrates Across Multiple Crystallographic Orientations}

	
	\author{Neethu Rose Leyo}
	\altaffiliation{These authors contributed equally to this work}
	\affiliation{Department of Physics $\&$ Astronomy, Louisiana State University, Baton Rouge,  LA 70803, USA}
	\author{Sethulakshmi Sajeev}
	\altaffiliation{These authors contributed equally to this work}
	\affiliation{Department of Physics $\&$ Astronomy, Louisiana State University, Baton Rouge,  LA 70803, USA}
	\author{Arnaud P. Nono Tchiomo}
	\altaffiliation{These authors contributed equally to this work}
	\affiliation{Department of Physics $\&$ Astronomy, Louisiana State University, Baton Rouge,  LA 70803, USA}
	\author{Prosper Ngabonziza}
	\email[corresponding author, ]{pngabonziza@lsu.edu}
	\affiliation{Department of Physics $\&$ Astronomy, Louisiana State University, Baton Rouge,  LA 70803, USA}
	\affiliation{Department of Physics, University of Johannesburg, P.O. Box 524 Auckland Park 2006, Johannesburg, South Africa}
	\date{\today}

	\begin{abstract}
		
 Atomically flat and chemically controlled oxide substrate surfaces are essential for high-quality epitaxial growth of complex materials and heterostructures. Previous studies have demonstrated \textit{in situ} thermal preparation of selected oxide substrates, primarily focusing on conventional low-index orientations such as SrTiO$_3$~(100) and LaAlO$_3$~(100). Here, we extend these methodologies to a broader range of oxide substrates and crystallographic orientations, including challenging polar surfaces. We present a systematic study of the \textit{in situ} thermal preparation of SrTiO$_3$~(100), SrTiO$_3$~(111), GdScO$_3$~(110), LaAlO$_3$~(111),  (LaAlO$_3$)$_{0.3}$(Sr$_2$AlTaO$_6$)$_{0.7}$~(001), and NdGaO$_3$~(110) single crystal substrates. The substrates are annealed in a vacuum chamber under a controlled oxygen background environment using a CO$_2$ laser heating system.  The optimized annealing conditions produce atomically flat step-and-terrace structures, long-range surface crystallinity, and effectively clean surfaces  across all the investigated substrates.  Our results establish a versatile, reproducible, and broadly applicable platform for the \textit{in situ} preparation of oxide substrates immediately prior to epitaxial growth.  These results also provide practical guidelines for interface engineering and surface-sensitive investigations of complex materials.

	\end{abstract}

	\maketitle

	\section{Introduction}

The physical properties of epitaxial thin films and heterostructures are strongly influenced by the structural and chemical quality of the surface of the underlying substrate, particularly at the film–substrate interface~\cite{biswas2017atomically,sanchez2014tailored}. Since epitaxial layers  nucleate and grow directly on the substrate, the initial surface termination, morphology, and crystallinity  play a critical role in determining the structural quality, defect density, strain state, and functional properties of the deposited materials~\cite{hwang2012emergent,mannhart2010oxide}. Precise control of the substrate's surface is therefore essential for achieving high-quality epitaxial growth and atomically engineered interfaces for applications ranging from electronics and spintronics to catalysis, energy materials, and quantum technologies~\cite{schlom2011interface,takagi2010emergent}.

 Complex oxides are attractive materials systems owing to their diverse electronic, magnetic, and ferroelectric properties, which can be tailored through strain and interface engineering~\cite{schlom2007strain,hwang2012emergent}. Perovskite oxide single crystal substrates such as SrTiO$_3$, LaAlO$_3$, NdGaO$_3$, GdScO$_3$, and (LaAlO$_{3}$)$_{0.3}$(Sr$_{2}$AlTaO$_6$)$_{0.7}$ (LSAT) are widely used for epitaxial growth because their lattice parameters and crystal symmetries enable controlled epitaxial strain engineering~\cite{sanchez2014tailored,haeni2004room}. Epitaxial strain can significantly influence electronic bandwidth, orbital occupation, magnetic ordering, metal–insulator transitions, and superconducting properties in complex oxide thin films and heterostructures \cite{chakhalian2012whither,boschker2017quantum,cen2008nanoscale}. However, to realize such strain- and interface-driven phenomena, atomically smooth and chemically controlled substrate surfaces are required to ensure reproducible high-quality epitaxy~\cite{kawasaki1994atomic}. Indeed, surface roughness, mixed termination, contamination, or structural disorder at the substrate level can lead to non-uniform film nucleation, increased defect formation in the film, interface intermixing, as well as  degraded thin film properties~\cite{koster1998quasi,rijnders2005build}.

 Conventional substrate preparation methods often rely on multi-step \textit{ex situ} chemical etching and high-temperature furnace annealing procedures to achieve atomically flat and single-terminated surfaces~\cite{kawasaki1994atomic,kleibeuker2010atomically,biswas2017atomically}. While these methods have been widely successful, it should be noted that they are frequently time-consuming, and involve hazardous chemicals such as buffered hydrofluoric acid (HF), a highly corrosive etchant. These \textit{ex situ} surface termination techniques also comprise the exposure of the substrates to air between preparation and deposition steps. This increases the risk of potential surface contamination, which would hamper the reproducibility of the method.  In addition, the \textit{ex situ} surface preparation becomes increasingly challenging for substrates with polar surfaces such as SrTiO$_3$~(111) and LaAlO$_3$~(111), where electrostatic instability, surface reconstruction, and step bunching can complicate the formation of atomically ordered surfaces~\cite{saghayezhian2016polar,woo2015surface}.
 
 An attractive alternative is the  \textit{in situ} thermal preparation of the substrates directly inside the deposition chamber using a laser-assisted annealing technique. This approach completely eliminates air exposure, thereby reducing surface contamination and providing a streamlined route for substrate preparation immediately prior to epitaxial deposition. It has been successfully demonstrated for selected oxide substrates with the preferential (001) crystallographic orientation, including SrTiO$_3$~(001), LaAlO$_3$~(001), NdGaO$_3$~(001), MgO~(001), and Al$_2$O$_3$~(0001)~\cite{jager2018independence,braun2020situ}. Building on the substrate preparation protocols described therein, this \textit{in situ} approach has subsequently been employed for the epitaxial growth of high-quality oxide thin films and heterostructures, demonstrating its compatibility with advanced oxide synthesis and interface engineering~\cite{tchiomo2022combined,ngabonziza2023apl,ngabonziza2024prm,wang2024interface,kim2025nitride,nono2019high,suceava2026colossal,sajeev2026electronic}. It is noteworthy that the applicability of the \textit{in situ} thermal preparation to oxide substrates with polar surfaces—which have also been found to be challenging to prepare using conventional \textit{ex situ} methods~\cite{chang2008atomically,russell2008surface}—remains unexplored.
 
In this work, we present a systematic study of the \textit{in situ} thermal preparation of SrTiO$_3$~(100), SrTiO$_3$~(111),  GdScO$_3$~(110), LaAlO$_3$~(111), LSAT~(001) and NdGaO$_3$~(110) single crystal substrates using a CO$_2$ laser heating system. Particular emphasis is placed on establishing a unified \textit{in situ} thermal preparation methodology applicable to oxide substrates with diverse crystal structures, surface polarities, and crystallographic orientations, while developing reproducible substrate-specific annealing protocols. This work provides the first systematic comparison of \textit{in situ}  thermal preparation across oxide substrates with multiple crystallographic orientations spanning both polar and non-polar surfaces. 
\begin{figure*}[!t]
    \centering
    \begin{minipage}[c]{0.70\textwidth}
        \centering
        \includegraphics[width=\linewidth]{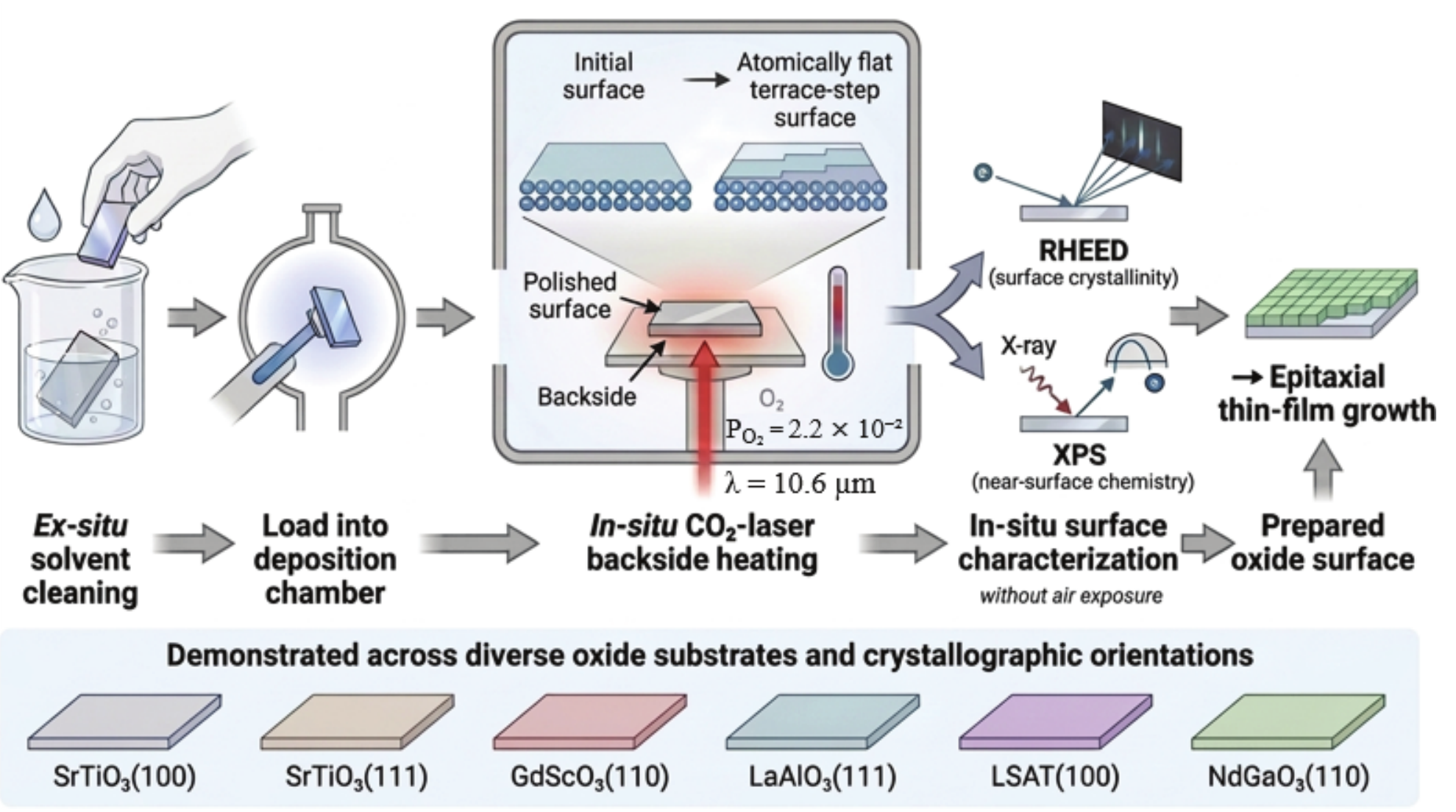}
    \end{minipage}
    \hfill
    \begin{minipage}[c]{0.27\textwidth}
        \caption{\label{fig:schematic}
       Schematic illustration of the surface preparation workflow, \textit{in situ} characterization, and range of substrates investigated in this work. The oxide substrates are heated from the backside using a CO$_2$ laser heating under a controlled oxygen pressure. The thermally treated surfaces are characterized \textit{in situ} by RHEED and XPS without being exposed to air. The approach is demonstrated across multiple oxide substrates and crystallographic orientations.}
    \end{minipage}
\end{figure*}

\section{Experimental details}\label{Experi}
	
Single-sided polished single crystal substrates, with dimensions 5$\times$5$\times$0.5 mm$^3$, were prepared using a two-step cleaning procedure. The procedure consisted of an \textit{ex situ} chemical cleaning process on the bench, followed by an \textit{in situ} thermal annealing inside a vacuum chamber. The \textit{ex situ} cleaning procedure was mainly  aimed at removing surface contaminants such as grease, dust particles, and organic residues. Initially, the polished side of each substrate was gently cleaned using a lint-free cloth wetted with isopropanol. The substrates were then sequentially ultrasonicated in separate glass beakers containing acetone, isopropanol, and deionized (DI) water for 10 min each, at room temperature. After each ultrasonication step, the substrates were transferred to the next solvent bath. Following the cleaning process, the substrates were dried using high-purity nitrogen gas. This standard cleaning procedure was employed for most substrates except LSAT~(001) and SrTiO$_3$~(111), for which modified cleaning conditions were used.

 For LSAT~(001), the substrates were only ultrasonicated in acetone and isopropanol, while the DI water cleaning step was intentionally omitted. This modification was motivated by the presence of droplet-like features at the surface of the substrates that were exposed to DI water. This morphology suggested that chemical alteration may have occurred at the surface prior to the \textit{in situ} thermal treatment, as discussed below. As for  SrTiO$_3$~(111), the \textit{ex situ} cleaning procedure differed slightly from the standard process described above.  After ultrasonic baths in acetone and isopropanol for 10~min each, the SrTiO$_3$~(111) substrates were further ultrasonicated in DI water at 80~$\degree $C for 30~min. This additional treatment was introduced to improve the surface preparation of the polar SrTiO$_3$~(111) surface. However, consistent with previous reports, we observed that prolonged exposure to DI water at elevated temperature can also lead to increased surface roughness~\cite{biswas2017atomically,saghayezhian2016polar}. 

After the \textit{ex situ} cleaning procedure, the substrates were mounted on a custom-designed sample holder comprising metal and ceramic components, compatible with an Omicron-type transfer system. Next, the substrates were loaded into the load lock and  subsequently transferred into the pulsed laser deposition  (PLD) chamber, where the \textit{in situ} thermal treatment was carried out. The annealing procedure was performed under a controlled oxygen background environment. Throughout the annealing process, the background pressure was maintained at 2.2$\times{10}^{-2}$~mbar. This pressure was found to provide stable and reproducible surface preparation conditions across all substrates investigated in this study. It is also comparable, in order of magnitude, to those employed in previous works on \textit{in situ} thermal preparation of oxide surfaces~\cite{jager2018independence,braun2020situ}.

The \textit{in situ} surface cleaning  was performed by directly heating the substrate from the backside using a CO$_2$ laser heating system operating at the  wavelength of 10.6~$\mu$m. The temperature was monitored and controlled using a built-in pyrometer directed at the unpolished backside of the substrate along the laser beam path. The pyrometer emissivity setting ($\varepsilon$) was set to 0.66, based on calibration experiments performed on an SrTiO$_3$~(100) substrate. For calibration, an 80~$\mu$m thick thermocouple (TC) wire was attached to the front surface of the substrate using a very small drop of silver paste, allowing simultaneous temperature measurements on the front and back surfaces. The temperature was increased in stages at a rate of 50~\degree C/min up to 1000~\degree C, and $\varepsilon$ was adjusted until the measured temperatures at both sides approached agreement. Analysis of the calibration data yielded $\varepsilon=0.63\pm0.06$ over the full temperature range. At the upper end of the range, the $\varepsilon$ value was found to be approximately 0.66, supporting the use of an emissivity setting of 0.66 for subsequent temperature control during annealing at higher temperatures. The uncertainty reflects the experimental limitations associated with TC contact and temperature measurements on small substrates.

The same emissivity setting was used for all substrate materials to maintain a consistent measurement configuration. However, because the calibration was performed only on SrTiO$_3$~(100) and did not extend to the maximum annealing temperatures, the reported temperatures should be interpreted as pyrometer-indicated values rather than independently verified absolute surface temperatures. Differences in the spectral emissivity and thermal response of the substrates may introduce material-dependent systematic errors. Nevertheless, the method was successfully applied to the different oxide substrates investigated in this work to identify annealing conditions that yielded well-ordered surface morphologies.  Furthermore, the annealing temperatures associated with the formation of well-defined, single-unit-cell-height terraces are broadly consistent with those reported in previous studies of in situ CO$_2$-laser thermal preparation of oxide surfaces~\cite{braun2020situ,jager2018independence}. This agreement provides additional support for the suitability of the selected annealing conditions, although it does not independently establish the accuracy of the absolute temperature measurements.  Accordingly, the reported temperatures primarily document the optimized processing conditions for each substrate and should not be interpreted as establishing a direct equivalence of absolute annealing temperatures across different materials.  The CO$_2$ laser heating system used for this work is supplied by SURFACE systems + technology GmbH \& Co. KG~\cite{surface2026}. 

For all substrates investigated, the annealing process consisted of four stages. These stages are indicated by the different slopes in the temperature profile shown in Fig.~\ref{STO-100}\textcolor{blue}{(a)}. In the first stage, the substrate's temperature is increased up to 600~$\degree$C at a rate of 100~$\degree$C/min. In the second stage, the temperature is further increased at a rate of 50~$\degree$C/min to reach the maximum annealing temperature. Under optimized conditions, this temperature corresponds to the temperature at which the surface is clean, atomically smooth, and has achieved well-ordered structures.  In the third stage, the substrate is maintained at the annealing temperature for a certain duration that was found to be substrate-dependent. In the last stage, the substrate is cooled to room temperature also at a rate of 50~$\degree$C/min. It should be highlighted that for NdGaO$_3$~(110), a modified cooling procedure was employed to overcome the cracking of the substrate that was observed during cooling. Following the annealing at the maximum temperature and for the selected dwell time, the substrate was first cooled down to 600~$\degree$C at a reduced rate of 25~$\degree$C/min, and subsequently cooled to room temperature at 50~$\degree$C/min.

During annealing, the surface crystallinity and ordering were monitored  using a TorrRHEED system  operated at the  accelerating voltage of 35~keV and the filament current of 1.5~A. The diffraction  patterns projected onto the fluorescent  screen were recorded using the kSA 400 software. Following the thermal treatment, the substrates  were transferred \textit{in situ} into the XPS  chamber without being exposed to air. XPS measurements were performed to determine the near-surface chemical composition of the substrates and assess their surface cleanliness. Although all the investigated substrates were characterized by XPS, only representative spectra for SrTiO$_3$~(100) and SrTiO$_3$~(111) are presented and discussed in this work. The XPS results for the remaining substrates will be reported in a separate study. Core level spectra of O~$1s$, Sr~$3p$, Sr~$3d$, and Ti~$2p$ were acquired using a monochromated Al~K$\alpha$ source and a  hemispherical energy analyzer (PHOIBOS 150). A quantitative analysis of the spectra through peak fitting helped to determine the relative surface stoichiometry and assess the near-surface termination. 

Owing to the electrically insulating nature of the substrates, an electron flood gun was employed for charge compensation during the measurements. The C~$1s$ signal acquired from the substrates  as-inserted was used as a reference for binding energy calibration. For subsequent XPS measurements on \textit{in situ} cleaned surfaces, the core level spectra acquired from the corresponding substrates as-inserted were used as references for energy calibration. This approach assumes that the optimized thermal treatment conditions do not induce chemical changes at the near-surface region of the substrates. In addition, the attenuation of the C~$1s$ signal was used as an indicator of the effectiveness of the \textit{in situ} cleaning procedure in removing surface contaminants.
 
After XPS characterization, the samples were removed from vacuum and their surface morphologies were analyzed \textit{ex situ }by AFM. For all substrates, the AFM measurements were performed in tapping mode at a scan rate of 1~Hz. Multiple regions across each substrate's surface were scanned to assess the lateral uniformity and homogeneity of the \textit{in situ} thermally prepared surfaces. Height profiles extracted from the AFM images were used to determine the step heights and characterize the surface flatness and smoothness.  The overall surface preparation workflow, \textit{in situ} characterization, as well as the range of oxide substrates  investigated in this study are schematically summarized in Fig.~\ref{fig:schematic}.
	 
 \section{Results and Discussion}
     \label{sec:results}

 In this section, we present the AFM and RHEED characterizations of the different substrates investigated. We also discuss the XPS data of representative substrates. The results are presented in  Sec.~\ref{sec:results}\,\ref{sec:sto100}--\ref{sec:ngo110} on a substrate-by-substrate basis, highlighting both common trends and distinct surface characteristics for each material system. The reproducibility of the surface morphology, crystallinity, and chemical cleanliness demonstrates the robustness and broad applicability of the \textit{in situ} thermal preparation method across multiple oxide substrates.
	 
\subsection{SrTiO$_3$~(100)}
\label{sec:sto100}
	
Given its extensive use in oxide epitaxy and its well-established surface preparation protocols~\cite{jager2018independence,kawasaki1994atomic,koster1998quasi,ohnishi2004preparation}, SrTiO$_3$~(100) was employed in this study as a benchmark substrate for validating the \textit{in situ} thermal preparation procedure and correlating the resulting surface morphology, crystallinity, and chemical composition.

\begin{figure}[!t]
	\includegraphics[width=0.48\textwidth]{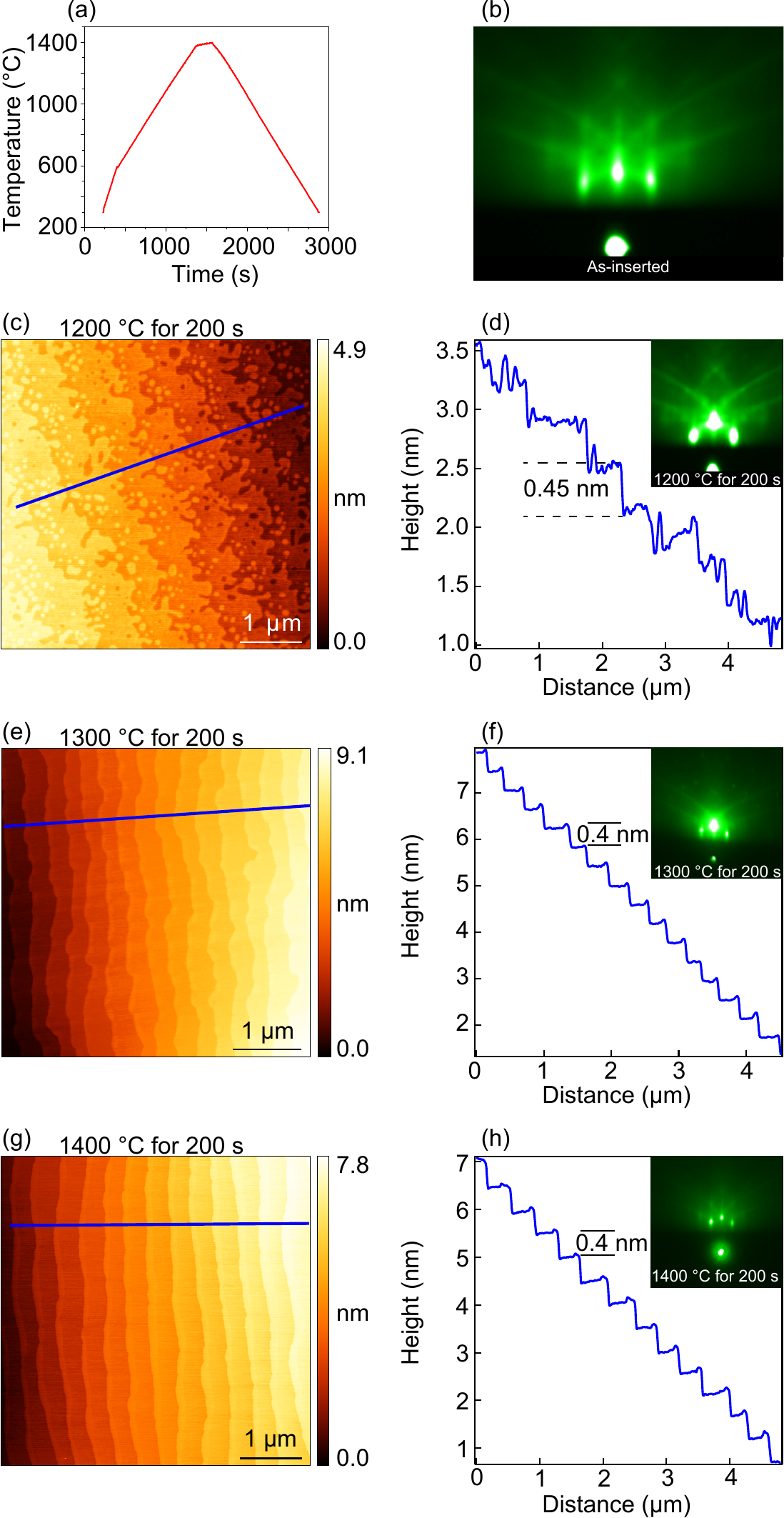}
	\caption{\textbf{\textit{In situ} thermal preparation and surface characterization
			of {SrTiO$_3$~(100)} substrates.} (a) Temperature profile of an annealing process, with the maximum annealing temperature reaching 1400~\degree C. (b) A RHEED image of the substrate as-inserted. AFM and RHEED images depicting the progression of the topography and structure of the surface upon annealing for 200~s at (c) and (d) 1200$\degree$C, (e) and (f) 1300$\degree$C, and (g) and (h) 1400$\degree$C. The height profiles on the right are extracted from the lines indicated on the corresponding AFM images on the left. The RHEED images are recorded at room temperature (RT).}		
	
	\label{STO-100}
\end{figure}

Figure~\ref{STO-100}\textcolor{blue}{(a)} shows a heating curve obtained during the \textit{in situ} thermal treatment of a SrTiO$_3$~(100) substrate. The annealing was conducted under a constant oxygen background pressure. The optimal annealing conditions were determined to be $1400~\degree$C and 200~s dwell time. This annealing temperature is consistent with that previously reported for the \textit{in situ}  annealing of the SrTiO$_3$~(100) substrates using a CO$_2$ laser~\cite{braun2020situ}.
	
Figures~\ref{STO-100}\textcolor{blue}{(b)}-\ref{STO-100}\textcolor{blue}{(h)} reveal the evolution of the surface morphology with increasing annealing temperatures. At the maximum annealing temperature of $1200~\degree$C [Fig.~\ref{STO-100}\textcolor{blue}{(c)}], terrace structures are already visible; however, numerous island-like features remain distributed across the surface with curved terrace edges, indicating incomplete surface reorganization. The corresponding height profile [Fig.~\ref{STO-100}\textcolor{blue}{(d)}] exhibits non-uniform step heights with weakly-resolved terrace edges. Upon increasing the annealing temperature to $1300~\degree$C [Fig.~\ref{STO-100}\textcolor{blue}{(e)}], the AFM micrograph displays a better resolved terrace structure. This is evident from the extracted height profile that reveals well-defined terraces with nearly straight edges and uniform step heights of 0.4~nm [see, Fig.~\ref{STO-100}\textcolor{blue}{(f)}].

\begin{figure*}[t]
	\centering
	\includegraphics[width=\textwidth]{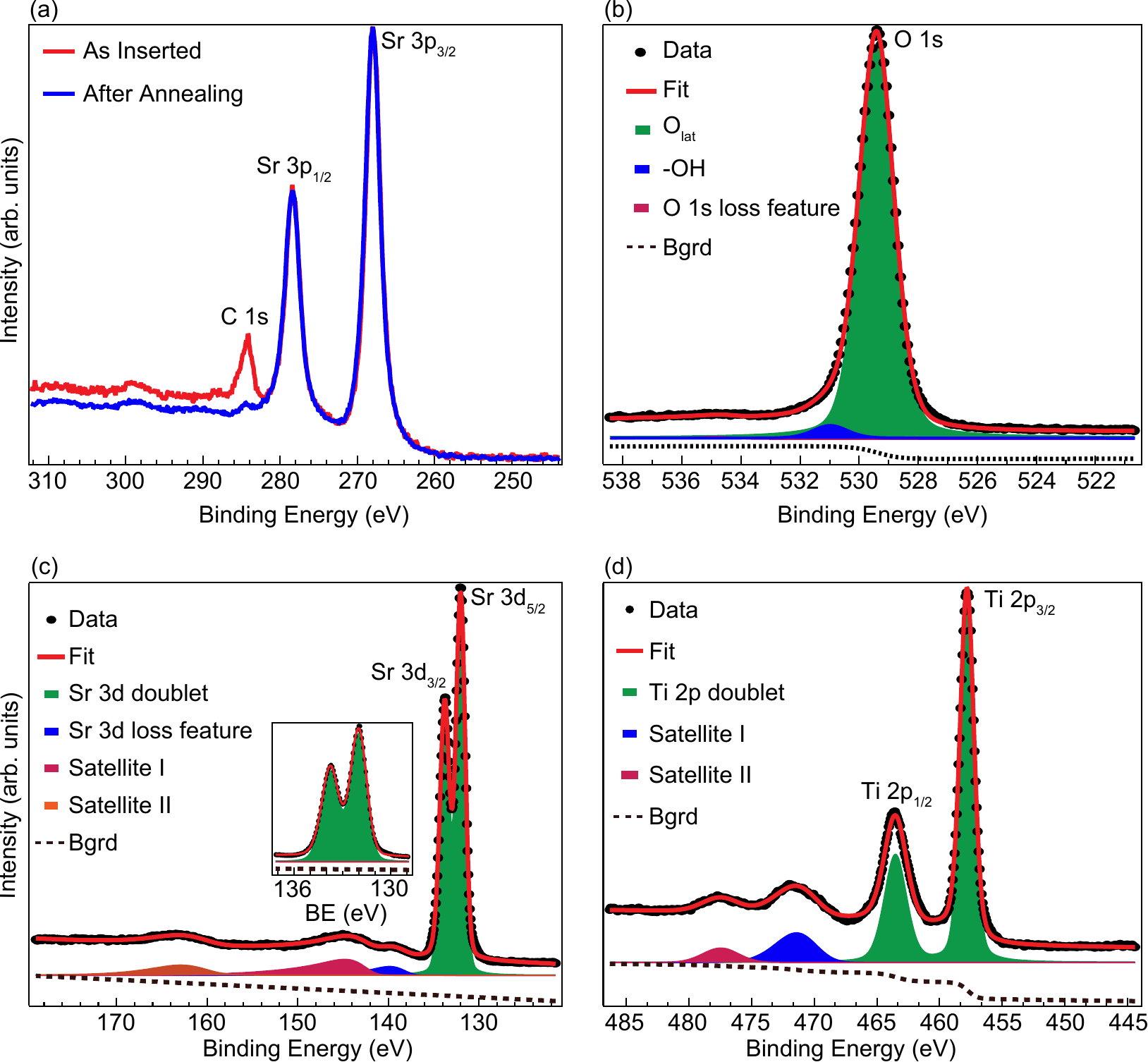}
	\caption{\textbf{High resolution scans of the XPS core level spectra of a SrTiO$_3$~(100) substrate following an \textit{in situ} thermal preparation at 1400~$\degree$C for 200~s}. (a) Sr~$3p$ spectra recorded before and after annealing, illustrating the removal of adventitious carbon contamination. Peak fitting of (b) O~$1s$, (c) Sr~$3d$, and (d) Ti~$2p$ obtained after annealing. The inset in (c) provides an enlarged view of the fitted Sr~$3d$ doublet peak.}
	\label{XPS-STO100}
\end{figure*}
 
When the annealing temperature is further increased to $1400~\degree$C [Fig.~\ref{STO-100}\textcolor{blue}{(g)}], an enhanced surface diffusion and improved surface ordering is achieved. This is demonstrated by the formation of atomically flat terraces with uniform widths and straighter edges separated by sharp and continuous steps [Fig.~\ref{STO-100}\textcolor{blue}{(h)}]. The measured step height of 0.4~nm  is consistent with the formation of single-unit-cell-height terraces, as it is closer to the room-temperature lattice parameter of SrTiO$_3$, $a = 3.905$~\AA~\cite{biswas2011universal,kawasaki1994atomic}. The terrace morphology originates from the slight substrate miscut introduced during substrate polishing and fabrication. During annealing, thermally activated surface diffusion promotes step-flow reorganization and surface free-energy minimization, thereby revealing the underlying vicinal structure of the substrate surface~\cite{kawasaki1994atomic,ohnishi2004preparation,biswas2017atomically}.

The RHEED pattern of the surface before annealing is shown in Fig.~\ref{STO-100}\textcolor{blue}{(b)}; and the insets in Figs.~\ref{STO-100}\textcolor{blue}{(d)}, \ref{STO-100}\textcolor{blue}{(f)} and \ref{STO-100}\textcolor{blue}{(h)} depict the RHEED patterns obtained after annealing the substrates at $1200~\degree$C, $1300~\degree$C and $1400~\degree$C for 200~s, respectively. Prior to annealing, the diffraction patterns already exhibit bright diffraction spots and relatively sharp streaks along with clear Kikuchi lines, indicating that the surfaces of as-inserted substrates possess a relatively well-ordered crystalline structure. These features provide a reference for assessing the evolution of surface crystallinity and ordering during subsequent thermal annealing. Following the thermal treatment and at the highest annealing temperature, the streaks become sharper and the diffraction spots smaller. This suggests an improved surface crystallinity associated with a well-ordered and smoother surface morphology, in good agreement with the AFM observations.

Figure~\ref{XPS-STO100} depicts the chemical composition of the surface of a SrTiO$_3$~(100) substrate.  Figure~\ref{XPS-STO100}\textcolor{blue}{(a)} compares the Sr~$3p$ core level spectra acquired before and after annealing at $1400~\degree$C. A significant reduction in the intensity of the C~$1s$ peak is observed in the spectrum of the annealed surface. This demonstrates the effectiveness of the \textit{in situ} thermal preparation in removing adventitious surface contamination. In addition, the line shape of the Sr~$3p$ peak remains essentially unchanged compared with that of the as-inserted surface, indicating that the annealing process preserves the chemical integrity of the surfaces while improving their cleanliness.

Fits of the O~$1s$, Sr~$3d$, and Ti~$2p$ core levels acquired from the surface annealed at 1400~\degree C are presented in Figs.~\ref{XPS-STO100}\textcolor{blue}{(b)}--\ref{XPS-STO100}\textcolor{blue}{(d)}. The O~$1s$ spectrum exhibits a nearly symmetric line shape and is well described by an intense component assigned to lattice oxygen (O$_{\mathrm{lat}}$), and a weaker component located at higher binding energy associated with hydroxyl groups [{Fig.~}\ref{XPS-STO100}\textcolor{blue}{(b)}]~\cite{larramona1989characterization,fujishima2008tio2,tchiomo2020electronic}. The absence of a pronounced higher-binding-energy component, together with the substantial attenuation of the C~$1s$ signal, is consistent with the effective removal of surface contaminants during \textit{in situ} thermal treatment~\cite{haasch2014sto001,ishida2008coherent,NonoTchiomo2018Surface}.

The Sr~$3d$ spectrum [Fig.~\ref{XPS-STO100}\textcolor{blue}{(c)}] is well described by the expected spin-orbit doublet, together with satellite and energy-loss features. The inset shows the excellent agreement between the experimental data and the fit. Similarly, the Ti~$2p$ spectrum [Fig.~\ref{XPS-STO100}\textcolor{blue}{(d)}] exhibits the characteristic Ti$^{4+}$ doublet without evidence of potential  chemical reduction or secondary phases~\cite{NonoTchiomo2018Surface}. 

A quantitative analysis of the fitted spectra was conducted to evaluate  the near-surface stoichiometry. The results are summarized in Table \ref{tab:sto100_composition}. The analysis reveals a near-stoichiometric Sr--Ti--O composition with a slight enhancement of the Sr signal relative to Ti. While this observation suggests a Sr-enriched near-surface region, conventional XPS probes several atomic layers and therefore does not provide an unambiguous determination of the outermost surface termination~\cite{raisch2012preparation,baeumer2015surface}. Nevertheless, these results demonstrate that the \textit{in situ} thermal preparation procedure produces a chemically clean and well-ordered SrTiO$_3$~(100) surface while preserving the overall Sr--Ti--O composition, in agreement with previous reports on thermally prepared SrTiO$_3$ substrates~\cite{kawasaki1994atomic,jager2018independence,braun2020situ,NonoTchiomo2018Surface}.

\begin{table}[t]
	\caption{Elemental composition of the \textit{in situ} annealed SrTiO$_3$~(100) surface determined from XPS measurements.}
	\label{tab:sto100_composition}
	\begin{ruledtabular}
		\begin{tabular}{lc}
			Element & Composition (\%) \\
			\hline
			Sr & 24.55 \\
			Ti & 16.80 \\
			O  & 58.65 \\
		\end{tabular}
	\end{ruledtabular}
\end{table}

\subsection{SrTiO$_3$~(111)}

{SrTiO$_3$~(111) consists of alternating stacks of charged Ti$^{4+}$ and [SrO$_3$]$^{4-}$ planes along the [111] direction, giving rise to a polar surface with an expected single-step height of approximately $2.25\,\text{\AA}$  \cite{biswas2017atomically,saghayezhian2016polar,woo2015surface}. However, the  resulting electrostatic instability often promotes surface reconstruction and step bunching, thereby complicating the preparation of a uniformly terminated surface.  This challenge has been reported for \textit{ex situ} surface preparation methods involving deionized water etching and chemical etching~\cite{woo2015surface,biswas2011universal,kumari2023realization}. Consequently, carefully optimized surface treatment conditions are required for the \textit{in situ} thermal preparation of the SrTiO$_3$~(111) substrates [see Section~\ref{Experi}]. Among the various combinations of maximum annealing temperature and annealing duration explored,  two yielded well-defined, atomically flat step-and-terrace structures: annealing at 1400~$\degree$C for 600~s and at 1500~$\degree$C for 220~s.
	
\begin{figure}[!t]
	\includegraphics[width=0.48\textwidth]{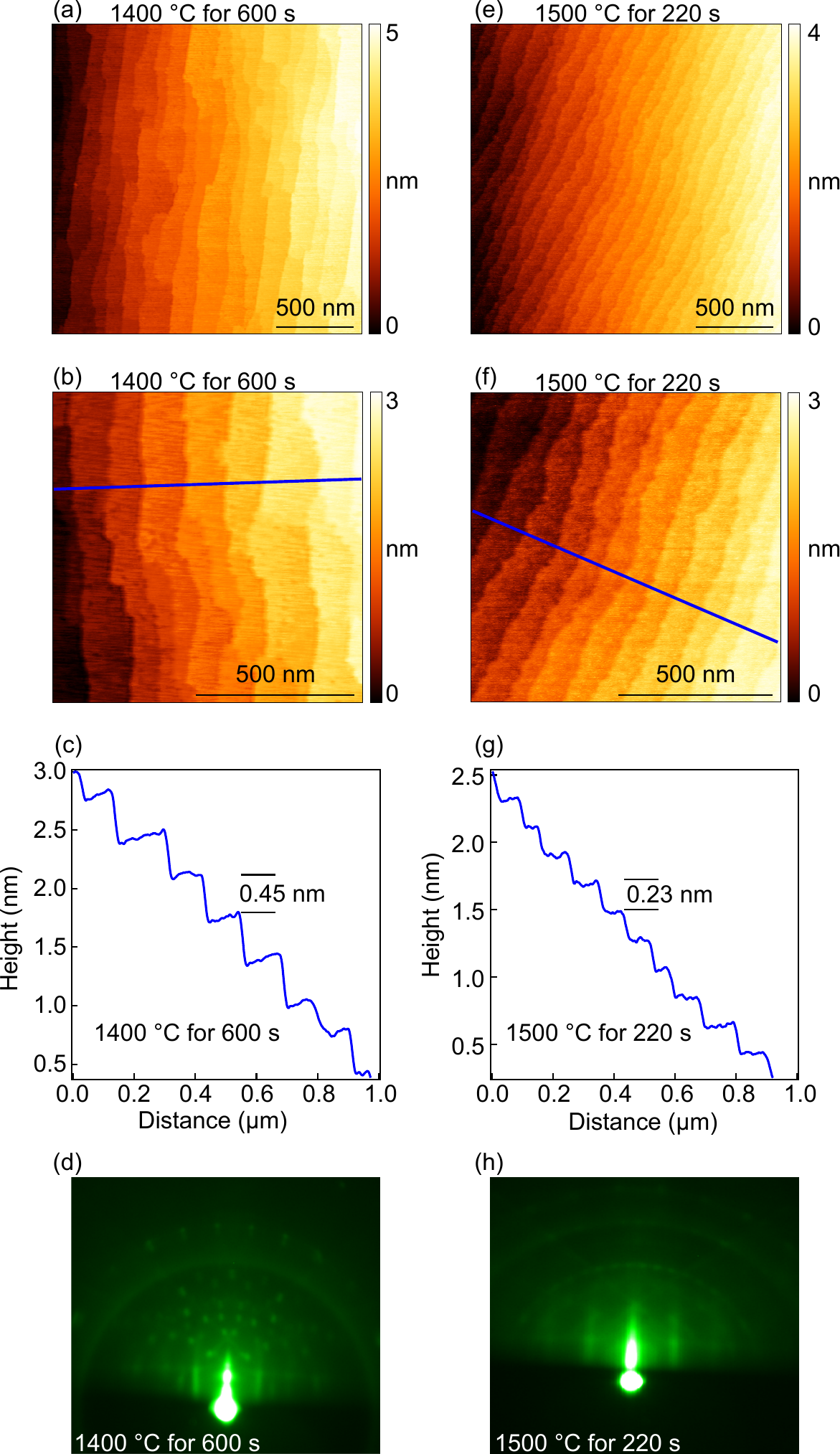}
	\caption{\textbf{\textit {In situ} thermal preparation and surface analysis of SrTiO$_3$~(111) substrates.} 
		Surface morphology and structure of the substrates annealed at (a)~--~(d) 1400~$\degree$C for 600~s, and at (e)~--~(h) 1500~$\degree$C for 220~s. The AFM images are given with different magnifications in (a) and (b), and in (e) and (f) as indicated by the scale bars. The corresponding height profiles in (c) and (g) are extracted from the lines marked on the images. The RHEED images in (d) and (h) are recorded after cooling the annealed substrates to RT.}
	
	\label{STO-111}
\end{figure}
	
 \begin{figure*}[t]
 	\centering
 	\includegraphics[width=\textwidth]{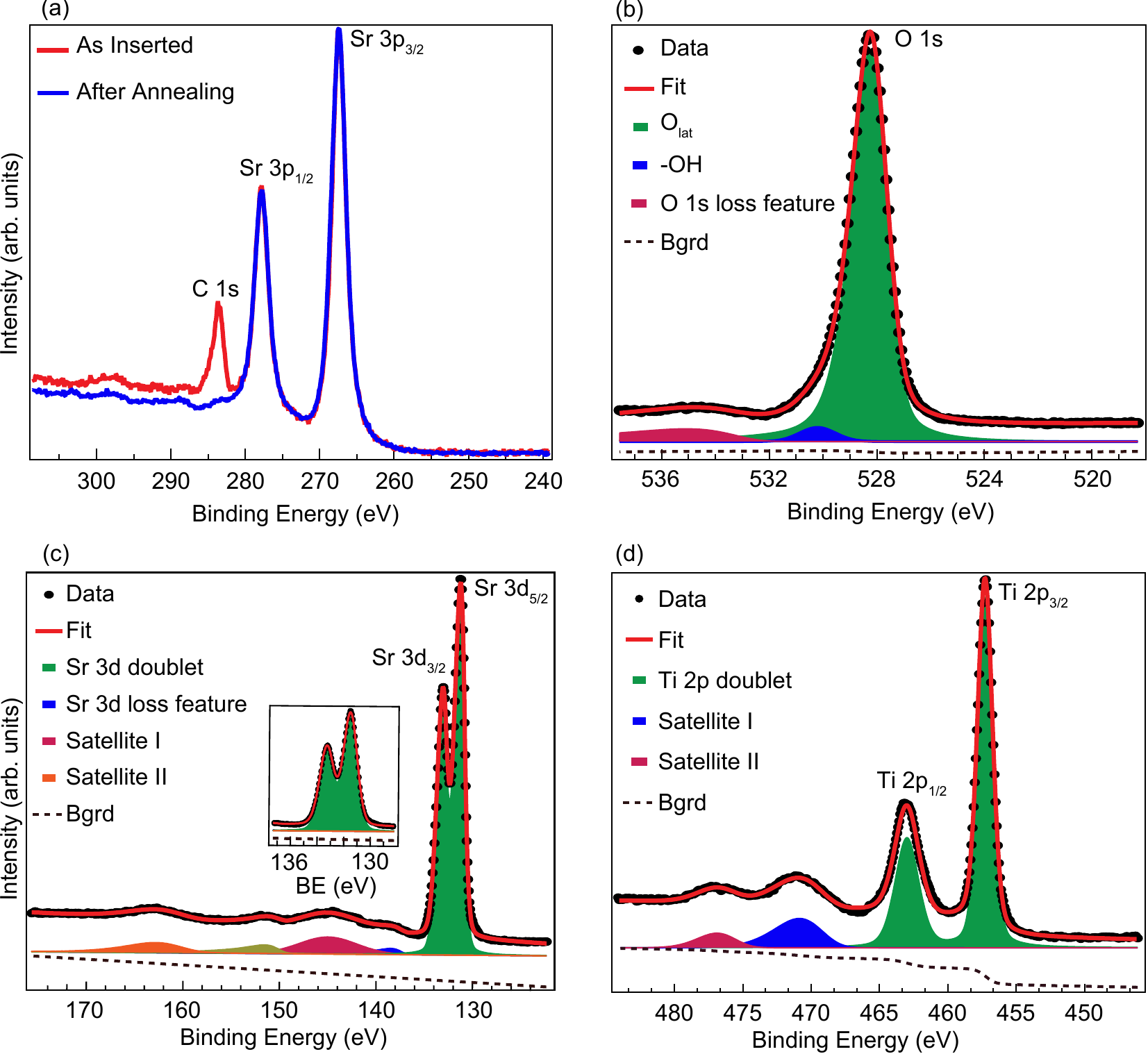}
 	\caption{\textbf{High resolution scans of the core levels of the SrTiO$_3$~(111) substrate annealed at 1500~\degree C for 220~s}. (a) Sr~$3p$ spectra before and after annealing. High-resolution scans of the core level spectra with fitted components for (b) O~$1s$, (c) Sr~$3d$, and (d) Ti~$2p$. The inset in (c) highlights the fitted Sr~$3d$ spin-orbit doublet.}
 	\label{XPS-STO111}
 \end{figure*}
 
Figure~\ref{STO-111} summarizes the \textit{in situ} thermal preparation and surface characterization of SrTiO$_3$~(111). For the surface annealed at 1400~\degree C for 600~s, the topography presents a terrace structure dominated by step bunching as shown in Fig.~\ref{STO-111}\textcolor{blue}{(a)}. A scan across a smaller region of this scanned area is presented in  Fig.~\ref{STO-111}\textcolor{blue}{(b)}. It shows some waviness on the edge of the terraces. Thus, locally, the terraces are large with fairly uniform width separated by uniform steps of approximately 0.45~nm [Fig.~\ref{STO-111}\textcolor{blue}{(c)}]. This value closely corresponds to twice the (111) interplanar spacing of SrTiO$_3$~(111), and could be associated with structural cohesion along the interplanar direction and the energetic gain of step bunching~\cite{bachelet2007step}. It is also consistent with steps connecting equivalent surface terminations, suggesting that significant step coalescence and surface restructuring remain under these annealing conditions~\cite{torrelles2019electronic,sanchez2014tailored}. The presence of fractional-order diffraction features observed in the RHEED image shown in Fig.~\ref{STO-111}\textcolor{blue}{(d)} tends to support the hypothesis of a distinct surface reconstruction. Nevertheless, the strong specular reflection along with the relatively sharp streak-like diffraction features corresponding to the bulk surface lattice support a smooth and reasonably ordered surface, consistent with the step-and-terrace structure observed in AFM. 

Upon increasing the annealing temperature to 1500~\degree C for 220~s, a significant improvement in surface morphology is achieved as shown  in Figs.~\ref{STO-111}\textcolor{blue}{(e)} and \ref{STO-111}\textcolor{blue}{(f)}. Here, the extended terraces are narrower with uniform width and straighter edges,  leading to a smoother, atomically flatter, and better-ordered surface. The corresponding height profile [Fig.~\ref{STO-111}\textcolor{blue}{(g)}] yields a regular step height of  about 0.23~nm. This is in good agreement with the expected SrTiO$_3$~(111) single-step height of $\sim$0.225~nm, equivalent to the spacing between adjacent Ti$^{4+}$ and SrO$_3^{4-}$ planes along the [111] direction~\cite{russell2008surface,biswas2017atomically,saghayezhian2016polar,woo2015surface,biswas2011universal}. The closer agreement with the expected step height, combined with the more sharply defined terrace morphology, suggests enhanced surface diffusion and more efficient step-flow reorganization at a higher annealing temperature. This is also evident in the RHEED image presented in Fig.~\ref{STO-111}\textcolor{blue}{(h)}, where only the integer-order diffraction patterns are visible with a significant sharpness, suggesting a well-defined long-range surface order.

The surface morphology of the \textit{in situ} prepared SrTiO$_3$~(111) substrate reported in this work is comparable to that achieved using the conventional technique which consists of  chemical etching followed by \textit{ex situ} annealing for several hours~\cite{woo2015surface,biswas2011universal,kumari2023realization}. This highlights the effectiveness and efficiency of our \textit{in situ} thermal preparation method, and, to the best of our knowledge, there is no experimental evidence of the \textit{in situ} preparation of the  SrTiO$_3$~(111) surface to this date.
 
 Figure~\ref{XPS-STO111} presents the XPS characterization of the surface of a SrTiO$_3$~(111) substrate. The Sr~$3p$ spectra acquired from the substrate as-inserted and following the thermal treatment at 1500~$\degree$C for 220~s are compared in Fig.~\ref{XPS-STO111}\textcolor{blue}{(a)}. Upon annealing, the C~$1s$ peak intensity is markedly attenuated, providing direct spectroscopic evidence that adventitious carbon contamination has been effectively eliminated from the surface. Concurrently, the Sr~$3p$ core levels exhibit essentially no change in line shapes. These are clear indications  that the annealing protocol not only preserves the intrinsic chemical environment, but it also  yields chemically clean and structurally intact substrate templates, suitable for subsequent epitaxial growth.

The residual C~$1s$ signal observed in the spectra of the annealed surfaces [Figs.~\ref{XPS-STO100}\textcolor{blue}{(a)} and~\ref{XPS-STO111}\textcolor{blue}{(a)}] is attributed to the gradual adsorption of carbon-containing species during XPS measurements. Despite maintaining a chamber pressure of approximately $1\times10^{-9}$~mbar throughout data acquisition, the C~$1s$ signal increased progressively over successive scans, indicating the gradual accumulation of carbon-containing contaminants on the initially clean surfaces. 

\begin{table}[t]
	\caption{Elemental composition of the \textit{in situ} annealed SrTiO$_3$~(111) surface determined from XPS measurements.}
	\label{tab:sto111_composition}
	\begin{ruledtabular}
		\begin{tabular}{lc}
			Element & Composition (\%) \\
			\hline
			Sr & 26.78 \\
			Ti & 17.37\\
			O  & 55.85\\
		\end{tabular}
	\end{ruledtabular}
\end{table}

Fits of the O~$1s$, Sr~$3d$, and Ti~$2p$ core level spectra acquired from the annealed surface are presented in Figs.~\ref{XPS-STO111}\textcolor{blue}{(b)}--\ref{XPS-STO111}\textcolor{blue}{(d)}. Again, the O~$1s$ spectrum is dominated by the lattice oxygen component (O$_{\mathrm{lat}}$) associated with oxygen in the SrTiO$_3$ crystal lattice. The weak component attributed to surface hydroxyl, $^-$OH, is also observed. The Sr~$3d$ spectrum is well reproduced by the expected spin-orbit doublet together with weak satellite and energy-loss components. The inset in Fig.~\ref{XPS-STO111}\textcolor{blue}{(c)} presents an enlarged view of the Sr~$3d$ doublet, demonstrating the excellent agreement between the experimental data and the fitted model in the main peak region. The relatively weak intensities of the loss features and satellites suggest that they primarily arise from intrinsic photoemission loss processes rather than additional Sr-containing chemical phases~\cite{baeumer2015surface,raisch2012preparation}. Likewise, the Ti~$2p$ spectrum is well described by the characteristic Ti$^{4+}$ spin-orbit doublet together with  satellite features [{Fig.~\ref{XPS-STO111}}\textcolor{blue}{(d)}]. No additional peaks attributable to lower-valence Ti species, such as Ti$^{3+}$ ~\cite{NonoTchiomo2018Surface}, are observed. Again, these results demonstrate the reliability of the \textit{in situ} annealing protocols that we have developed [see Fig.~\ref{fig:S3} of the supplemental materials].

A quantitative analysis of the fitted spectra provides additional insight into the near-surface composition of the thermally prepared SrTiO$_3$~(111) substrate. The surface elemental concentration is summarized in Table~\ref{tab:sto111_composition}. The measured Sr signal is enhanced relative to Ti. However,  as outlined above, the XPS measurements represent an average over several atomic layers and therefore cannot uniquely identify the outermost surface termination~\cite{raisch2012preparation,baeumer2015surface}. Nevertheless, this quantitative assessment is particularly important for the polar SrTiO$_3$~(111) surface, where high-temperature annealing may induce surface reconstruction, cation redistribution, and polarity-compensation mechanisms~\cite{torrelles2019electronic,saghayezhian2016polar,woo2015surface}.

	\subsection{GdScO$_3$~(110)}
	
GdScO$_3$ has an  orthorhombic perovskite crystal structure consisting of alternating GdO$^{+}$ and ScO$_2^{-}$ layers. It is characterized by a wide bandgap, high dielectric permittivity, and excellent mechanical stability~\cite{taherian2026visible}. Pseudocubic GdScO$_3$~(110) has a relatively large lattice parameter, 3.966~\text{\AA}, which enables substantial tensile strain in a wide range of material systems~\cite{Uecker2008Properties}. These properties make GdScO$_3$ an attractive platform for the epitaxial growth of functional materials and heterostructures~\cite{Park2020TwinfreeEL,Kan2013ThicknessDependentSR,schubert2003structural}. Here, we demonstrate, for the first time, the \textit{in situ} thermal preparation of GdScO$_3$~(110) substrates, yielding surfaces suitable for subsequent thin film deposition.
	
\begin{figure}[!t]
	\centering
	\includegraphics[width=\linewidth]{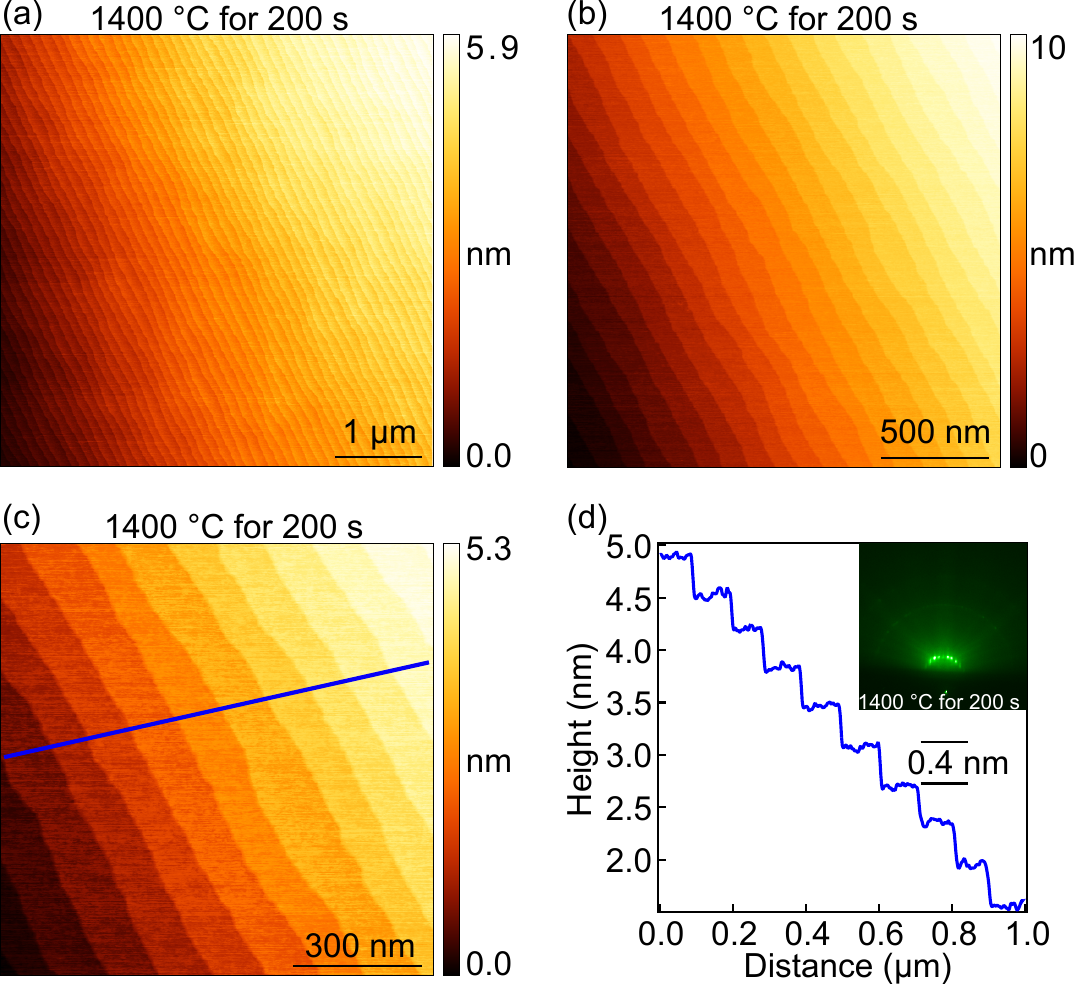}
	\caption{\textbf{\textit{In situ} thermal preparation and surface characterization of a GdScO$_3$~(110) substrate}. (a) - (c) AFM topography of the substrate's surface annealed at 1400~$\degree$C for 200~s. The surface morphology is shown at various magnifications as indicated by the scale bars. (d) Height profile extracted from the line scan indicated in (c).  The inset represents the RHEED pattern of the annealed surface, recorded at RT.}
	\label{GSO-110}
\end{figure}
		
Figure~\ref{GSO-110} depicts the surface characteristics of a GdScO$_3$~(110) substrate annealed at an optimal temperature of 1400~$\degree$C for the duration of 200~s. The AFM topography images shown in Figs.~\ref{GSO-110}\textcolor{blue}{(a)}--\ref{GSO-110}\textcolor{blue}{(c)} reveal extended step-and-terrace structures with laterally continuous terraces and well-defined step edges. The terrace widths are very narrow and somewhat uniform; the edges exhibit a slight meandering morphology, which may originate from anisotropic surface diffusion and step-edge relaxation during high-temperature annealing~\cite{bachelet2007step}. The height profile displayed in Fig.~\ref{GSO-110}\textcolor{blue}{(d)} exhibits well-resolved terrace edges, as well as straight and uniform steps of about 0.4~nm. The measured step height closely matches the pseudocubic unit-cell height of GdScO$_3$~(110) ($\sim 0.396$ nm), suggesting that the thermal treatment resulted in terraces separated by a single-unit-cell step~\cite{biswas2017atomically}. Furthermore, the RHEED pattern characterized by sharp and bright diffraction spots  on the zeroth Laue zone indicates long-range ordering of the surface.  Our \textit{in situ} thermal treatment protocol has yielded an atomically flat and smooth GdScO$_3$~(110) surface, similar to that realized with the \textit{ex situ} annealing method, thereby providing an excellent template for epitaxial oxide growth~\cite{mansley2020structure,biswas2017atomically}.

\subsection{LaAlO$_3$~(111)}

The perovskite substrate LaAlO$_3$ is widely used for oxide epitaxy, owing to its pseudocubic lattice parameter of approximately 3.79~\text{\AA}, and its ability to impose significant compressive strain on a variety of material systems~\cite{liu2012atomically,schlom2007strain}. Along the [111] direction, the crystal structure consists of alternating charged ({LaO}$_3)^{3-}$ and $\text{Al}^{3+}$ atomic planes, resulting in a highly polar surface. Such polarity can promote surface reconstruction, mixed termination, and other compensation mechanisms that influence the nucleation and interfacial structure of subsequently deposited films~\cite{middey2012epitaxial}. Here, following the \textit{in situ} annealing, we demonstrate the in situ thermal preparation of atomically flat LaAlO$_3$~(111) surfaces.

The AFM topography images of a surface annealed at 1400~$\degree$C for 200~s are presented in Figs.~\ref{LAO-111}\textcolor{blue}{(a)}--\ref{LAO-111}\textcolor{blue}{(c)}. The images reveal extended step-and-terrace structures with well-defined, continuous step edges, particularly evident in the 15$\times$15~$\mu$m$^2$ scan shown in Fig.~\ref{LAO-111}\textcolor{blue}{(a)} and the corresponding magnification across a 5$\times$5~$\mu$m$^2$ region in Fig.~\ref{LAO-111}\textcolor{blue}{(b)}. The terraces are wide, although the step edges exhibit waviness over both micrometer and shorter distance scales. Interestingly, as shown in Fig.~\ref{LAO-111}\textcolor{blue}{(c)}, even wider terraces are observed from a 5$\times$5~$\mu$m$^2$ scan of a different location on the surface. This variation in width may arise from local differences in step spacing induced by the substrate miscut angles, along with the rearrangement of surface steps during high-temperature annealing~\cite{yao2016surface}. Nonetheless, over large distances, the terraces are very uniform and the step heights are regular, as evidenced in the height profiles associated with both scanned regions and shown in Figs.~\ref{LAO-111}\textcolor{blue}{(d)} and \ref{LAO-111}\textcolor{blue}{(e)}. Both profiles exhibit step heights of 0.23~nm, denoting consistent step-height characteristics across different scan areas. The obtained value is in excellent agreement with the expected LaAlO$_3$~(111) interplanar spacing of 0.219~nm~\cite{herranz2012high,xu2016misfit}, calculated from the pseudocubic lattice parameter of LaAlO$_3$, through $ a_p = 3.79$\,\AA\, using the relation $d_{111}=a_p/\sqrt{3}$~\cite{koirala2018rich}. The close agreement between the measured and expected step heights is consistent with the formation of nearly single-layer terraces.

\begin{figure}[!h]
	\centering
	\includegraphics[width=\linewidth]{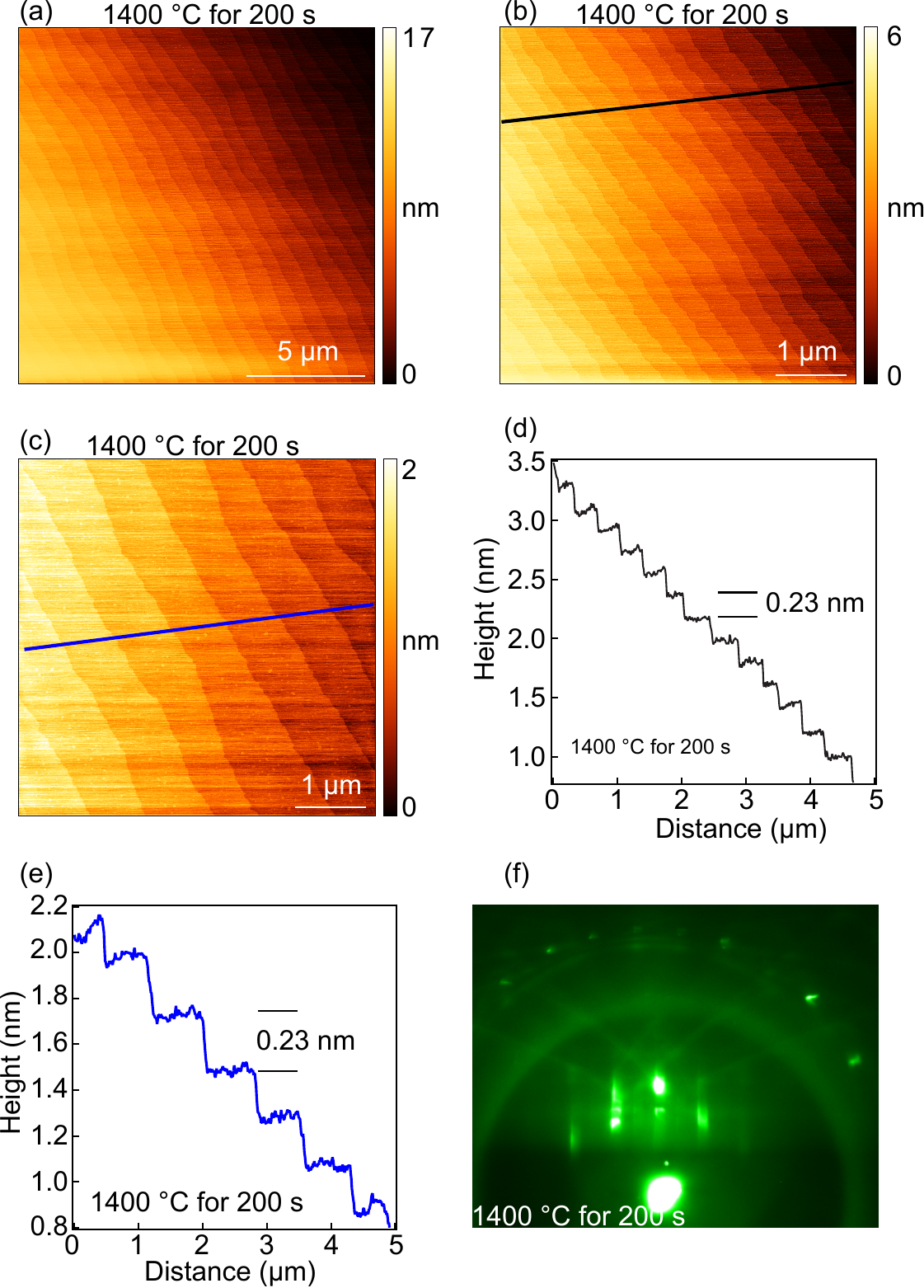}
	
	\caption {{\textbf{\textit{In situ} thermal preparation and surface characterization of a LaAlO$_3$~(111) substrate}. (a) - (c) AFM morphology of the substrate annealed at 1400~$\degree$C for 200~s. The image in (b) is a magnification across a smaller region of the area scanned in (a); and the image in (c) is a scan of a different location on the substrate's surface. (d) and (e) Height profiles extracted along the lines marked in (b) and (c), respectively. (f) RHEED image of the annealed surface recorded at RT.} }
	\label{LAO-111}
\end{figure}

Figure~\ref{LAO-111}\textcolor{blue}{(f)} shows the RHEED pattern recorded at RT after annealing. The pattern shows sharp and intense streak-like diffraction features, accompanied by distinctive Kikuchi lines. This reveals enhanced surface crystallinity and long-range ordering, consistent with the AFM results. The combined AFM and RHEED measurements therefore confirm that \textit{in situ} thermal preparation at 1400~\degree C for 200~s produces an atomically ordered LaAlO$_3$~(111) surface suitable for subsequent epitaxial growth. To the best of our knowledge, we believe that these results represent the first report of the \textit{in situ} thermal preparation of the surface of a LaAlO$_3$~(111) substrate.

\subsection{LSAT~(001)}

LSAT is a cubic perovskite substrate with a lattice parameter of approximately 3.868~\text{\AA}, providing an excellent close lattice match to SrTiO$_3$ and many other functional materials~\cite{ngai2010achieving}. Owing to this favorable lattice matching, LSAT has become a popular substrate for the epitaxial growth of high-temperature superconducting films and other complex material systems~\cite{ogawa2002fabrication,kuma2019enhancement}. In addition, LSAT remains electrically insulating under reducing conditions, making it particularly attractive for transport studies of epitaxial thin films~\cite{chakoumakos1998thermal,ngai2010achieving}. Unlike simple perovskites such as SrTiO$_3$, however, LSAT is chemically more complex, with mixed cation occupancy on both the A-sites (La/Sr) and B-sites (Al/Ta). This mixed-cation nature complicates the control of surface termination and makes conventional chemical etching with subsequent \textit{ex situ} annealing procedures less straightforward~\cite{biswas2017atomically}. Hence, as presented in Section~\ref{Experi}, we developed a careful preparation methodology for the \textit{in situ} annealing of the LSAT~(001) substrates to achieve surfaces suitable for subsequent epitaxy.

\begin{figure}[t!]
	\centering
	\includegraphics[width=0.46\textwidth]{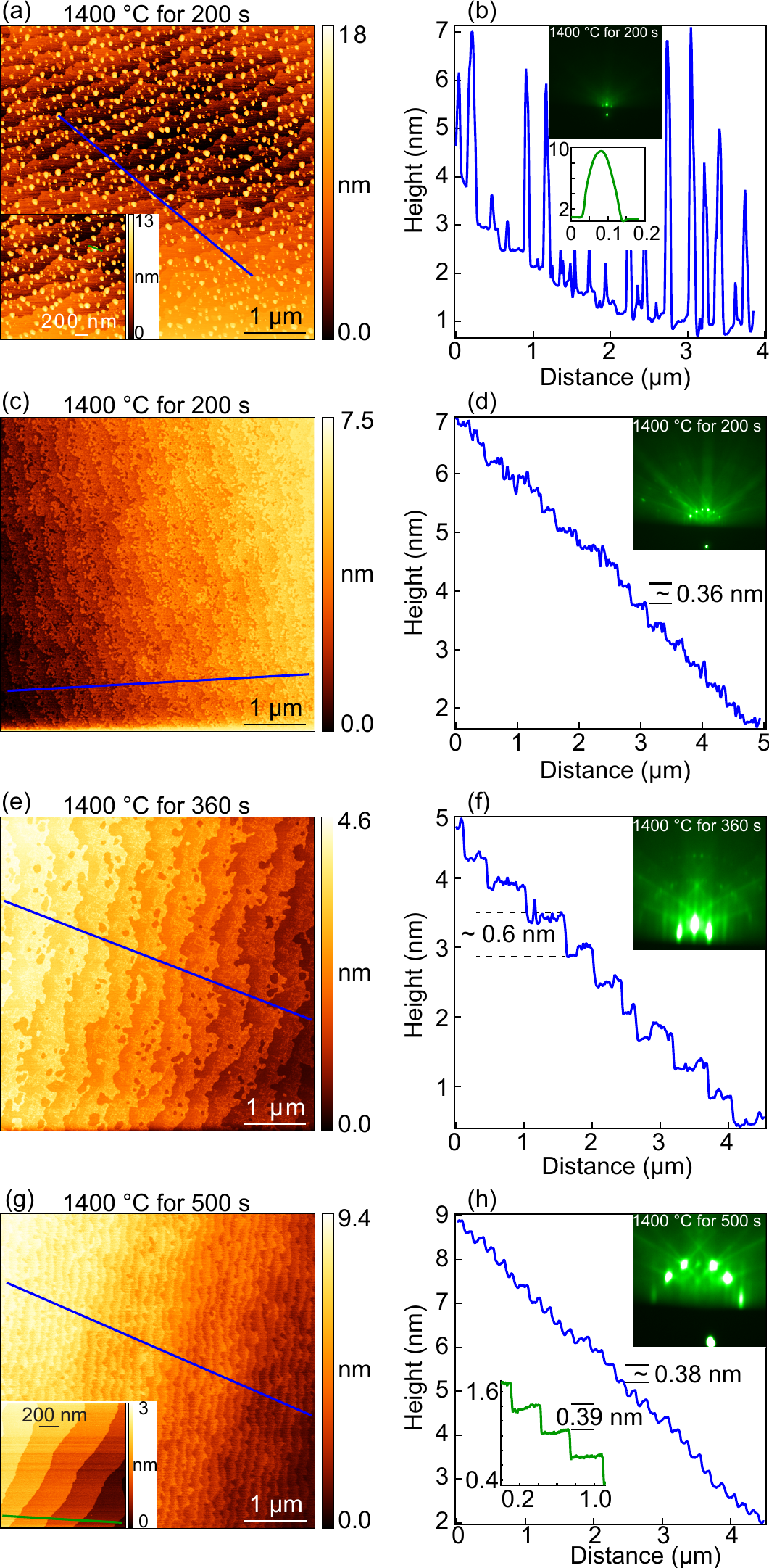}
	 \caption{{\textbf{\textit{In situ} thermal preparation and surface characterization of LSAT~(001) substrates}. AFM and RHEED data describing the evolution of the morphology and structure of the surface as a function of the annealing duration at 1400~$\degree$C. The annealing times are (a) - (d) 200~s, (e) and (f) 360~s, and (g) and (h) 500~s. In (a) and (b), the substrate was \textit{ex situ} cleaned in DI water. This step was omitted during the \textit{ex situ} cleaning of the surfaces shown in (c) - (h). The inset in (a) is a scan across a smaller region of the area in (a). The inset in (g) represents the scan of a smaller area at a different location on the surface in (g). The height profiles on the right are extracted from the lines indicated on the corresponding AFM images on the left. The bottom insets in (b) and (h) correspond to the surface profiles of the AFM insets in (a) and (g). The RHEED images are recorded at RT. }}
	\label{LSAT-100}
\end{figure}

Several annealing conditions were investigated, with the best surface quality obtained for a substrate annealed at 1400~\degree C for 500~s. As noted in Section~\ref{Experi}, achieving this surface quality required omitting the \textit{ex situ} bench cleaning step involving ultrasonic treatment in DI water. 

Figures~\ref{LSAT-100}\textcolor{blue}{(a)}--\ref{LSAT-100}\textcolor{blue}{(d)} compare the surface morphologies of two substrates subjected to identical \textit{in situ} annealing conditions, namely a maximum temperature of 1400~\degree C for 200~s. A comparison of the  surface initially cleaned in DI water [Figs.~\ref{LSAT-100}\textcolor{blue}{(a)} and \ref{LSAT-100}\textcolor{blue}{(b)}] with the surface that was not exposed to DI water [Figs.~\ref{LSAT-100}\textcolor{blue}{(c)} and \ref{LSAT-100}\textcolor{blue}{(d)}] reveals a marked difference in surface morphology. On the former, the developing terrace structure is largely obscured by rounded features resembling three-dimensional islands, whereas those features are absent on the latter. These surface mounds reach heights of up to 10~nm and lateral widths of up to 84~nm, as determined from the surface profile shown in the inset in Fig.~\ref{LSAT-100}\textcolor{blue}{(b)}.  Similar surface structures (mounds~$+$~terrace) were previously observed at the surfaces of LSAT substrates annealed \textit{ex situ} in air for several hours at temperatures ranging from 650~\degree C to 1300~\degree C~\cite{pradeep2022lsat,ngai2010achieving,ohashi2014surface,tomar2018realization}. 

Previous studies have reported that the mounds can be removed from LSAT surfaces by prolonged ultrasonication of the annealed substrates in DI water~\cite{pradeep2022lsat,tomar2018realization}. Based on this observation, the mounds were attributed to SrO, which is thought to accumulate at the surface as the A-plane, comprising Sr, La, and O, partially decomposes at high temperatures. This process has been attributed to La evaporation from the surface and/or diffusion into the bulk leading to SrO clustering at surface~\cite{pradeep2022lsat,ngai2010achieving,tomar2018realization}. Several approaches have been reported to control or eliminate these features. For instance, annealing LSAT substrates in air in the vicinity of La rich compounds, such as LaAlO$_3$, was found to reduce the density and size of the mounds~\cite{pradeep2022lsat}. Their complete elimination was achieved by combining a flowing O$_2$ gas with an increased La vapor pressure during annealing~\cite{ngai2010achieving}.  

Similar SrO mounds coexisting with step-and-terrace structures have also been observed at the surfaces of SrTiO$_3$~(001) substrates annealed at temperatures around 950~\degree C and 1100~\degree C~\cite{kumari2023realization,szot2000formation}. In these systems, the formation of mounds was associated with the agglomeration of surface particulates at elevated temperatures, where increased atomic mobility promotes surface restructuring~\cite{szot2000formation}.

Considering that exposure to DI water may alter the near-surface chemistry of LSAT, particularly in the presence of a Sr-rich termination, we hypothesize that the surface of the DI water-exposed substrate underwent chemical modification prior to annealing. As in the case of SrTiO$_3$~\cite{szot2000formation}, such modification may favor the formation of droplet-like features that subsequently agglomerate in the course of heat treatment. With prolonged annealing, these features may evolve into larger crystallites with SrO-rich surfaces, consistent with the surface morphology presented in Fig.~\ref{LSAT-100}\textcolor{blue}{(a)}. Again, looking at the surface morphology of the substrate not soaked in DI water [Fig.~\ref{LSAT-100}\textcolor{blue}{(c)}] but \textit{in situ} annealed under the same conditions, a terraced structure without droplet-like features can be clearly seen. 

This important result suggests a twofold observation. First, the formation of SrO mounds appears to be associated with the prior exposure of the substrate to DI water, rather than being an inevitable consequence of a thermal treatment alone. This interpretation suggests that the initial surface chemistry may play a critical role in mound formation. Although previous studies have attributed SrO mound formation to La loss through evaporation or diffusion into the bulk~\cite{pradeep2022lsat,ngai2010achieving,tomar2018realization}, our observations point to an additional possible mechanism involving chemical modification during \textit{ex situ} bench cleaning. This interpretation is consistent with the report of no detectable A-site decomposition or chemical change in the surface termination ratio after annealing the LSAT substrate at 1300~\degree C for 72~h~\cite{bachelet2011conducted}. Second, unlike the SrO mounds reported in previous studies, which can be difficult to remove and may require specialized annealing conditions involving La-containing compounds~\cite{pradeep2022lsat,ngai2010achieving}, the mounds observed here can be avoided by appropriately controlling the \textit{ex situ} cleaning procedure and subsequent \textit{in situ} thermal treatment.

The surface characteristics presented in Figs.~\ref{LSAT-100}\textcolor{blue}{(c)}--\ref{LSAT-100}\textcolor{blue}{(h)} are those of substrates \textit{ex situ} cleaned while omitting the DI water bath step. These figures portray the evolution of the LSAT surface as a function of the annealing duration at 1400~\degree C.  Annealing temperature and duration are interdependent, with the annealing time duration playing a critical role in the topographical condition of LSAT~\cite{pradeep2022lsat,ngai2010achieving}. As the annealing duration increases from 200~s to 500~s, the surface morphology evolves from a terrace structure that is weakly developed, broad, with strong meandering edges and 2D-islands in Figs.~\ref{LSAT-100}\textcolor{blue}{(c)} and \ref{LSAT-100}\textcolor{blue}{(e)}, to a well-developed and atomically flat  step-and-terrace morphology in Fig.~\ref{LSAT-100}\textcolor{blue}{(g)}. The surface morphologies portrayed in Figs.~\ref{LSAT-100}\textcolor{blue}{(c)} and \ref{LSAT-100}\textcolor{blue}{(e)} are consistent with those realized upon annealing in tube furnaces at 1050~\degree C for 1~h in O$_2$  gas pressure~\cite{boschker2011uniaxial}, and at 1100~\degree C and 1200~\degree C for 2~h  in air~\cite{bachelet2011conducted}. As for the surface structure displayed in Fig.~\ref{LSAT-100}\textcolor{blue}{(g)}, although the terrace edges still meander on the long-$\mu$m scale~\cite{ngai2010achieving,boschker2011uniaxial,bachelet2011conducted}, a scan across a smaller area at a different location on the surface depicts parallel terraces separated by continuous step edges extending laterally across the entire scanned area [see inset of Fig.~\ref{LSAT-100}\textcolor{blue}{(g)}]. The corresponding height profiles plotted in Fig.~\ref{LSAT-100}\textcolor{blue}{(h)} yield fairly regular steps on the multi-$\mu$m scale, but regular and straight steps on the short scale. Both measured step heights are comparable, and their value is around 0.39~nm, in excellent accord with one unit cell of LSAT ($\sim 3.868\,$\AA) \cite{nunley2016optical,li2003ordering}. 

Further insights into the evolution of the surface's structure with the annealing duration are provided by the corresponding RHEED patterns illustrated in the insets in Figs.~\ref{LSAT-100}\textcolor{blue}{(b)}, \ref{LSAT-100}\textcolor{blue}{(d)}, \ref{LSAT-100}\textcolor{blue}{(f)}, and \ref{LSAT-100}\textcolor{blue}{(h)}. At the dwell time of 200~s, the DI-water-treated surface shows relatively weak diffraction spots, with diffuse and less intense Kikuchi features [Fig.~\ref{LSAT-100}\textcolor{blue}{(b)}]. This surface structure is most probably due to the presence of the SrO mounds at the surface. In comparison, when the DI water treatment is not involved, the surface of the substrate annealed for 200~s exhibits brighter diffraction spots with well-resolved streaks and more distinguished Kikuchi lines [Figs.~\ref{LSAT-100}\textcolor{blue}{(d)}]. Following a thermal treatment with a higher dwell time of 360~s, the diffraction pattern becomes significantly brighter and the Kikuchi lines become more clearly defined. With a dwell time of 500~s, the diffraction features become sharper and remain brighter, and the Kikuchi lines become even more defined, indicating an improved surface ordering with prolonged annealing. Together with the AFM data, these observations demonstrate a well-ordered surface with a high crystallinity that could be employed for subsequent epitaxy. 

Our controlled and carefully optimized surface preparation technique has yielded a surface morphology consisting of terraces separated by a single-unit-cell step. This surface structure is similar to that obtained by \textit{ex situ} annealing in tubular furnace at comparable temperatures~\cite{ngai2010achieving,tomar2018realization,bachelet2011conducted}. However, the timescales necessary to achieve such atomically flat surface should be highlighted. With our well-developed \textit{in situ} annealing method, only a couple of minutes are needed, while several hours are required to obtain the same result using the \textit{ex situ} cleaning method~\cite{ngai2010achieving,tomar2018realization,bachelet2011conducted}. To the best of our knowledge, this study represents the first report of the \textit{in situ }thermal preparation of LSAT(001) substrates.  Regarding the chemical nature of the surface termination, previous studies have predominantly reported A-site terminated surfaces~\cite{ngai2010achieving}. However, determining the surface termination of the \textit{in situ} annealed LSAT substrates is beyond the scope of the present study, and will be addressed in a separate, dedicated investigation.

\subsection{NdGaO$_3$~(110)}
\label{sec:ngo110}

NdGaO$_3$ is an orthorhombically distorted perovskite, which is extensively used as substrate for epitaxial growth of complex oxide thin films. At room temperature, NdGaO$_3$ crystallizes in an orthorhombic GdFeO$_3$-type structure with $Pbnm$ symmetry~\cite{leca2012termination}. For a 110-oriented NdGaO$_3$ substrate, the corresponding pseudocubic (pc) unit cell is slightly distorted from the ideal cubic perovskite structure with lattice parameters of approximately $a{\mathrm{_{pc}}} = b{\mathrm{_{pc}}} = 3.863$~\AA\ and $c_{\mathrm{pc}} = 3.854$~\AA~\cite{vasylechko2000crystal,leca2012termination}.

\begin{figure}[!t]
	\includegraphics[width=\linewidth]{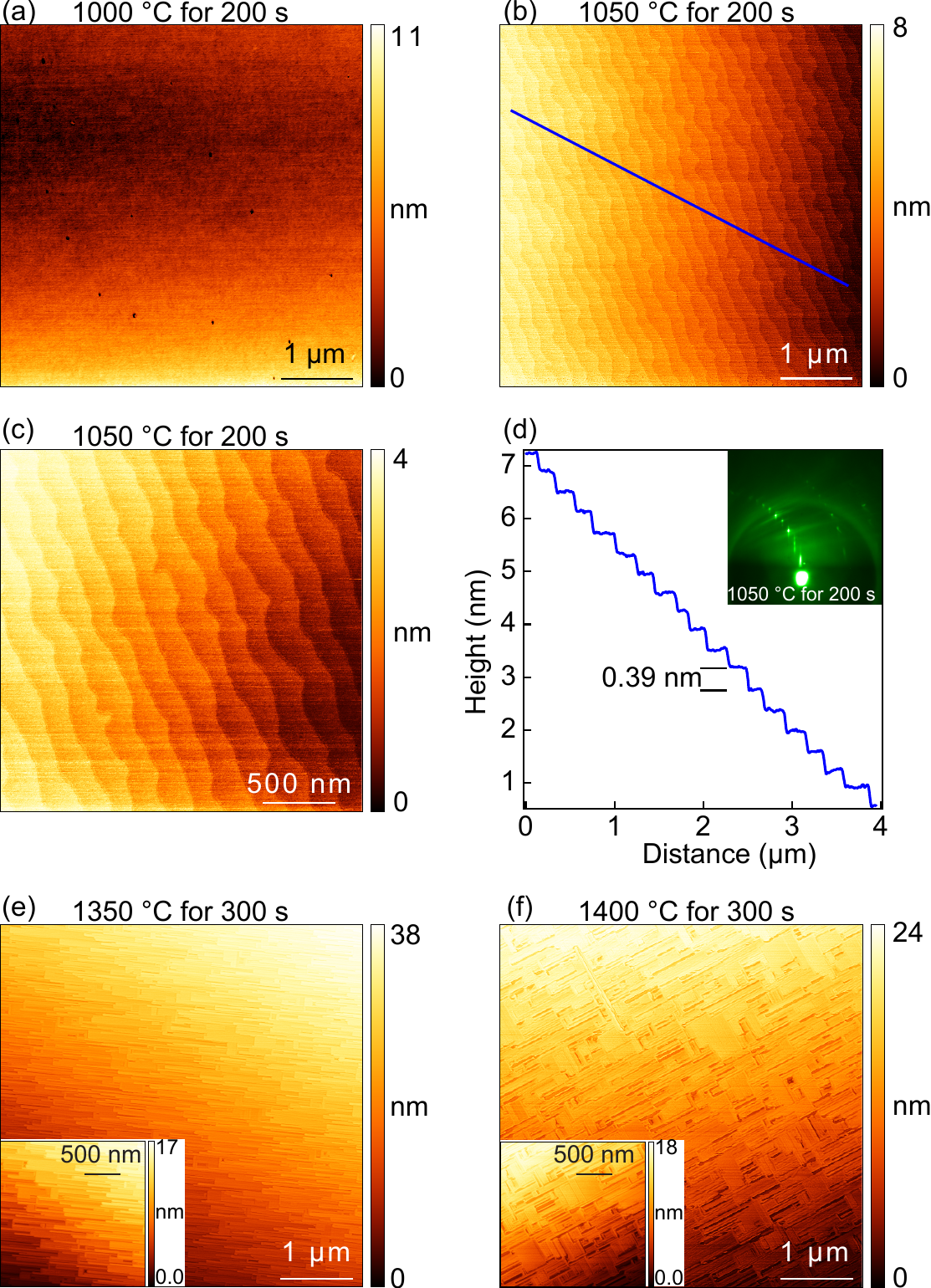}
	\caption{\textbf{\textit{In situ} thermal preparation and surface analysis of NdGaO$_3$~(110) substrates}. AFM data portraying the change in surface topography following the annealing of the substrates at (a) 1000~$\degree$C for 200~s, (b) - (d) 1050~$\degree$C for 200~s, (e) 1350~$\degree$C for 300~s, and (f) 1400~$\degree$C for 300~s. The image in (c) is a higher magnification of a region of the area scanned in (b). The height profile in (d) is extracted along the line indicated in (b); and the inset shows the RT RHEED pattern of the surface after an anneal at 1050~$\degree$C for 200~s. The insets in (e) and (f) are higher magnifications of regions of the respective scanned areas.}
	
	\label{NGO-110}
\end{figure}

\begin{table*}[t!]
	\centering
	\caption{Summary of the optimized annealing conditions for the oxide single crystal substrates investigated.}
	\label{tab:summary}
	\begin{tabular}{lcccccc}
		\toprule
		Substrate & Orientation & T (\degree C) & Time (s) & Step Height (nm) & Expected (nm) & Remarks \\
		\midrule
		SrTiO$_3$   & (100) & 1400 & 200 & 0.40 & 0.390     \cite{biswas2011universal,kawasaki1994atomic}  & {Sr-enriched near-surface composition } \\
		SrTiO$_3$   & (111) & 1500 & 220 & 0.23 & 0.225 \cite{saghayezhian2016polar,woo2015surface} & Polar surface \\
		GdScO$_3$   & (110) & 1400 & 200 & 0.40 & 0.396 \cite{biswas2017atomically,kleibeuker2010atomically} & Single-unit-cell steps \\
		LaAlO$_3$   & (111) & 1400 & 200 & 0.23 & 0.219 \cite{herranz2012high,xu2016misfit} & Polar surface \\
		LSAT  & (001) & 1400 & 500 & 0.39 & 0.387 \cite{nunley2016optical,li2003ordering} & Mixed-cation substrate \\
		NdGaO$_3$  & (110) & 1050 & 200 & 0.39& 0.386
		\cite{leca2012termination,zhou2023ferroelectricity} & orthorhombically distorted perovskite \\
		\bottomrule
	\end{tabular}
\end{table*}

For the \textit{in situ} thermal preparation of the NdGaO$_3$~(110) substrate,  the modified cooling procedure described in Section~\ref{Experi} was employed to minimize the risk of substrate cracking. Figure~\ref{NGO-110} depicts the gradual change in the surface morphology of NdGaO$_3$~(110) substrates under various annealing conditions. As shown in Fig.~\ref{NGO-110}\textcolor{blue}{(a)}, annealing at 1000~$\degree$C for 200~s did not produce clearly defined terraces, and the surface remained relatively featureless. However, when the annealing temperature is slightly elevated to 1050~$\degree$C for the same annealing time period of 200~s [Figs.~\ref{NGO-110}\textcolor{blue}{(b)} and \ref{NGO-110}\textcolor{blue}{(c)}], more prominent terrace features become visible, indicating a considerable improvement in surface morphology. The terraces are well-resolved with wavy edges extending continuously over several $\mu$m. The extracted height profile shown in Fig.~\ref{NGO-110}\textcolor{blue}{(d)} exhibits well-defined and uniform steps, with a height of 0.39~nm. This value is in good agreement with the expected pseudocubic lattice parameter of NdGaO$_3$~(110) ($\sim 3.86\,$\AA)~\cite{leca2012termination,zhou2023ferroelectricity,leca2011improved}, indicating the formation of single-unit-cell high terraces subsequent to the heat treatment. The corresponding RHEED pattern displayed in the inset reveals distinct and clearly resolved diffraction streaks on the Laue circle. These features indicate excellent crystalline ordering of the NdGaO$_3$~(110) surface at the optimized annealing temperature of 1050~$\degree$C, consistent with the atomically flat terrace morphology observed by AFM. 

For all the substrates investigated above, atomically flat and smooth surfaces were achieved at annealing temperatures around 1400~\degree C. This motivated us to investigate the surface morphology of NdGaO$_3$~(110) substrates annealed at temperatures exceeding the optimized annealing temperature of 1050~\degree C. Figures~\ref{NGO-110}\textcolor{blue}{(e)} and \ref{NGO-110}\textcolor{blue}{(f)} depict the surface morphologies of the substrates annealed for a duration of 300~s at 1350~$\degree$C and 1400~\degree C, respectively. Initially, the surface structure presents closely-spaced and elongated rectangular-like terraces with varying lengths and widths, predominantly aligned along a common direction [Fig.~\ref{NGO-110}\textcolor{blue}{(e)}].  Upon slightly ramping up the annealing temperature, the surface morphology becomes more complex, characterized by irregularly arranged and intersecting features with nonuniform dimensions [Fig.~\ref{NGO-110}\textcolor{blue}{(f)}]. These observations suggest that increasing the annealing temperature beyond the optimized 1050~\degree C modifies the surface evolution, potentially through enhanced surface diffusion and the development of step bunching or faceting.  Additionally, given the close proximity of these annealing temperatures to the melting temperature of NdGaO$_3$~(110)~\cite{ngmelting}, the observed morphologies may also reflect surface degradation associated with changes in surface stoichiometry or defect populations. Although similar morphologies were previously reported for NdGaO$_3$~(001) substrates \textit{in situ} annealed at 1095--1147~$\degree$C under an O$_2$ pressure of  $7.5\times 10^{-2}$~mbar~\cite{braun2020situ}, to the best of our knowledge, the present study provides the first report of such morphologies on NdGaO$_3$~(110). 

 \section{Conclusions}

Using a CO$_2$ laser heating system, we have demonstrated  systematic \textit{in situ} thermal preparation procedures for a broad range of oxide single crystal substrates with different crystallographic orientations, including SrTiO$_3$~(100), SrTiO$_3$~(111), GdScO$_3$~(110), LaAlO$_3$~(111), LSAT~(001) and NdGaO$_3$~(110). A summary of the optimized annealing parameters is provided in Table \ref{tab:summary}. The resulting surfaces exhibited  atomically flat step-and-terrace  morphologies, as confirmed by AFM, along with enhanced surface crystallinity and long-range ordering, as verified by RHEED. The surface cleanliness of the annealed substrates was further assessed \textit{in situ} by XPS. The optimized preparation conditions varied across substrate systems, reflecting differences in crystal symmetry, surface polarity, and surface chemistry. Reproducible surface preparation across all investigated substrates demonstrates the  versatility and robustness of the approach. Furthermore, AFM measurements acquired at different locations on the same substrate revealed consistent morphologies, confirming the spatial homogeneity of the prepared surfaces.

This \textit{in situ} process eliminates the need for hazardous HF-based chemical etching, prevents air exposure between surface preparation and thin film growth, substantially limits surface contamination, and significantly shortens the overall substrate preparation workflow~\cite{jager2018independence,braun2020situ}. Moreover, because the thermal treatment is performed directly within the deposition chamber, it integrates seamlessly with PLD, MBE, and thermal laser epitaxy (TLE)  systems~\cite{jager2018independence,kraay2023surface,braun2019film,smart2021thermal,kim2023thermal}, enabling immediate epitaxial deposition on atomically flat, chemically clean, and well-defined surfaces. The present work further broadens the applicability of this approach to a diverse range of oxide substrates, including polar surfaces, while establishing optimized annealing protocols for each substrate. Collectively, these results provide a versatile platform for oxide substrate preparation and practical guidelines for reliable epitaxy, interface engineering, oxide heterostructure fabrication, and surface-sensitive studies, thereby supporting the development of advanced oxide electronic devices across a broad range of complex oxide systems~\cite{ngabonziza2021advmat,kim2025high,suceava2026colossal}.

\section{Acknowledgments}

P.N. acknowledges startup funding from the College of Science and the Department of Physics $\&$ Astronomy at Louisiana State University. The authors gratefully acknowledge Dr. Philip Sprunger for providing the electron flood gun, which was essential for the XPS measurements of the insulating oxide substrates.
\bibliography{references}

\clearpage
\onecolumngrid
\newpage
\setcounter{table}{0}
\setcounter{figure}{0}
\renewcommand{\thefigure}{S\arabic{figure}}%
\setcounter{equation}{0}
\renewcommand{\theequation}{S\arabic{equation}}%
\setstretch{1.5}
\begin{center}
	\title*{\textbf{\large{Supplementary Information:} \\ [0.25in] \large{\textit{In situ} Thermal Preparation by CO$_2$-Laser and Chemical Characterization of Oxide Substrates Across Multiple Crystallographic Orientations}}}
\end{center}
\begin{center}
\normalsize{Neethu Rose Leyo$^{1,*}$, Sethulakshmi Sajeev$^{1,*}$, Arnaud P. Nono Tchiomo$^{1,*}$,\\[3pt] \hspace{80pt}and Prosper Ngabonziza$^{1,2,\ddagger}$}
\newline 
\\
{\footnotesize\textit{$^{1}$Department of Physics \& Astronomy, Louisiana State University, Baton Rouge, Louisiana 70803, USA
\newline $^{2}$Department of Physics, University of Johannesburg, P.O. Box 524, Auckland Park 2006, Johannesburg, South Africa}}

\end{center}
\onecolumngrid


%
%

\begin{center}
	\textbf{\large{Additional XPS Results}}
\end{center}

\begin{center}
\textbf{\large{a. XPS Survey Spectra}}
\end{center}

Fig.~\ref{fig:S2} presents the XPS survey spectra of the thermally prepared STO (100) and STO (111) substrates following \textit{in situ} CO$_2$-laser annealing. Both spectra exhibit the characteristic core-level peaks of Sr, Ti, and O, confirming the expected elemental composition of the substrates. Only weak carbon-related features are observed, indicating minimal surface contamination after thermal preparation and \textit{in situ} transfer to the XPS chamber. No additional impurity peaks are detected within the measurement sensitivity, demonstrating the chemical cleanliness of the prepared surfaces prior to AFM characterization and subsequent thin-film growth.

\begin{figure*}[!h]
	\centering
	\includegraphics[width=1\textwidth]{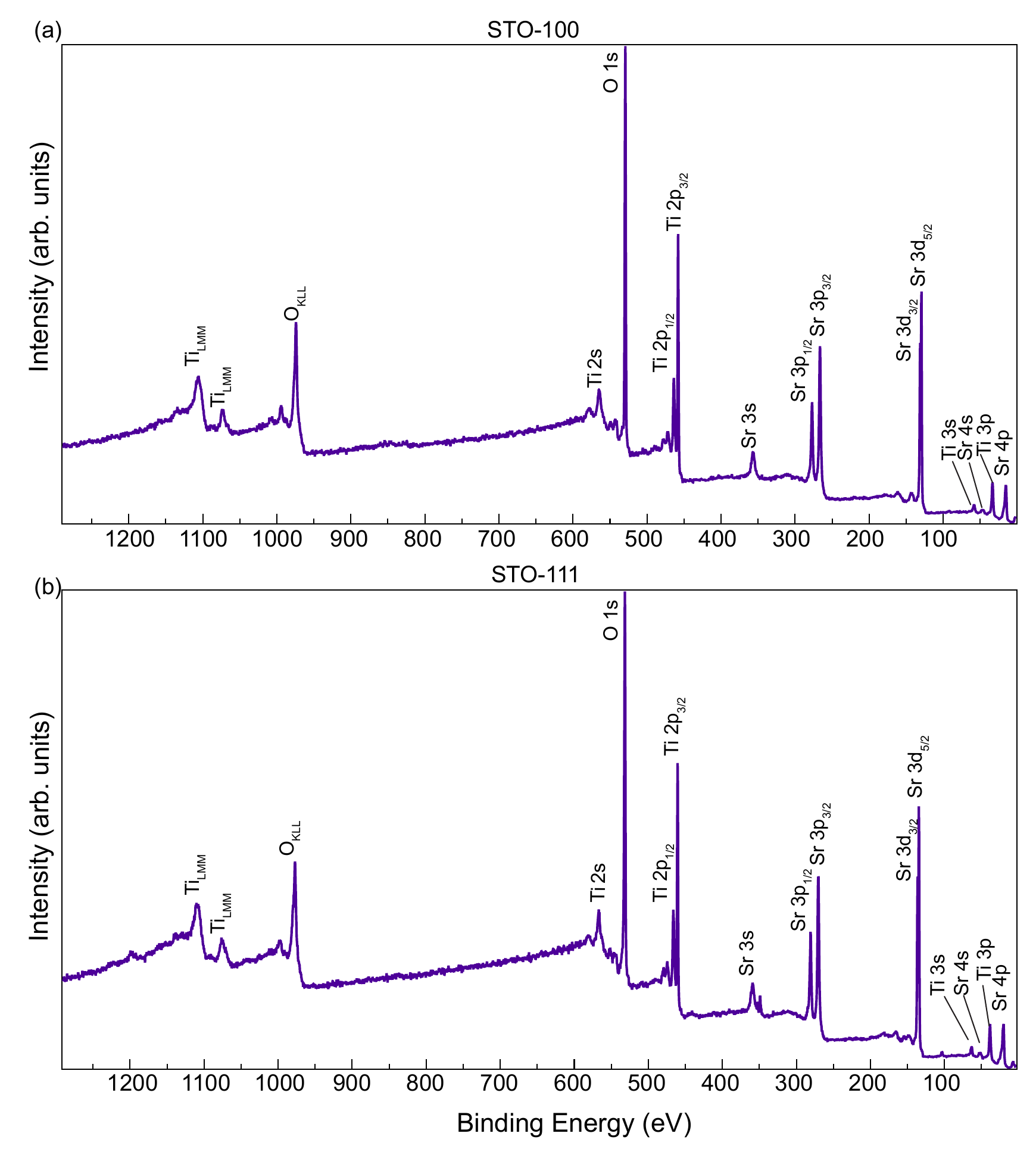}
	\caption{XPS survey spectra of the thermally prepared (a) STO(100) and (b) STO (111) substrates acquired after \textit{in situ} CO$_2$-laser annealing. The spectra show the characteristic Sr, Ti, and O core-level peaks with only weak carbon-related features, confirming the chemical cleanliness of the prepared surfaces and the absence of detectable impurity species.}
	\label{fig:S2}
\end{figure*}

\begin{center}
	\textbf{\large{b. XPS core-level Spectra}}
\end{center}

 The O 1s, Sr 3d, and Ti 2p core-level spectra were compared before and after in situ annealing, as shown in Fig.~\ref{fig:S3}. Fig.~\ref{fig:S3}\textcolor{blue}{(a)}-\ref{fig:S3}\textcolor{blue}{(c)} present the O 1s, Sr 3d, and Ti 2p spectra, respectively, for SrTiO$_3$ (100), while Fig.~\ref{fig:S3}\textcolor{blue}{(d)}-\ref{fig:S3}\textcolor{blue}{(f)} show the corresponding core-level spectra for SrTiO$_3$ (111). For both orientations, the higher binding-energy contribution in the O 1s spectra, associated with surface contamination or adsorbed species, decreases after annealing, indicating a cleaner surface following the thermal treatment. In contrast, the Sr 3d and Ti 2p core-level spectra show no pronounced changes in their main spectral features after annealing, suggesting that the chemical states of Sr and Ti are largely preserved.

\begin{figure*}[!h]
	\centering
	\includegraphics[width=1\textwidth]{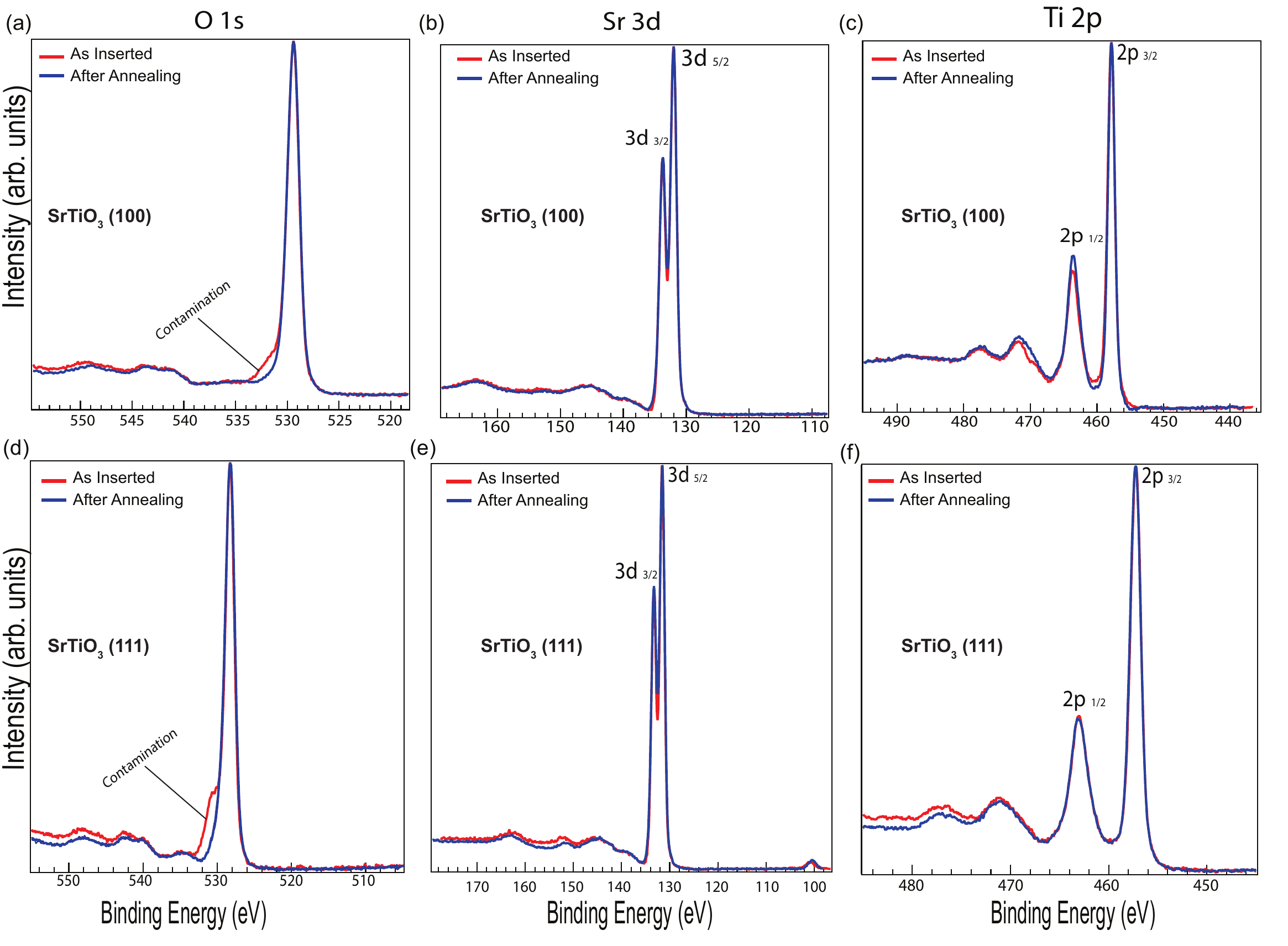}
	\caption{XPS core-level spectra of SrTiO$_3$ before and after \textit{in situ} annealing: (a) O 1s, (b) Sr 3d, and (c) Ti 2p spectra of SrTiO$_3$ (100); (d) O 1s, (e) Sr 3d, and (f) Ti 2p spectra of SrTiO$_3$ (111).}
	\label{fig:S3}
\end{figure*}
	
\end{document}